\documentclass[11pt, letterpaper]{article}

\usepackage{mathtools}
\usepackage{times}
\usepackage{bm}

\DeclarePairedDelimiter{\abs}{\lvert}{\rvert}
\newcommand{\Iacc}{\mathop{I}\nolimits_\mathrm{acc}}

\usepackage{ifthen}

\newcommand{\ifndef}[2]{\ifthenelse{\isundefined{#1}}{#2}{}}

\newcommand{\mydef}[2]{\def#1{#2}}

\newcommand{\nospell}[1]{#1}   %

\usepackage{amssymb}
\usepackage{amsmath}   %
\usepackage{amsthm}   %
\usepackage{latexsym}
\usepackage{amsfonts}
\usepackage{psfrag}    %
\usepackage{url}    %

\usepackage[dvips]{graphicx}
\usepackage[usenames,dvipsnames]{color}

\ifndef{\theorem}{\newtheorem{theorem}{Theorem}[section]}
\ifndef{\lemma}{\newtheorem{lemma}[theorem]{Lemma}}
\ifndef{\corollary}{\newtheorem{corollary}[theorem]{Corollary}}
\ifndef{\remark}{\theoremstyle{remark} \newtheorem{remark}{Remark}}
\ifndef{\proposition}{\newtheorem{proposition}{Proposition}}
\newtheoremstyle{mydefinition}   %
{\topsep}{\topsep}   %
{\slshape}   %
{}   %
{\bfseries}   %
{.}   %
{ }   %
{}   %
{\theoremstyle{mydefinition}}
\newtheoremstyle{myexample}   %
{\topsep}{\topsep}   %
{\itshape}   %
{}   %
{\slshape}   %
{:}   %
{ }   %
{\ul{\thmname{#1}}}   %
\ifndef{\example}{\theoremstyle{myexample} \newtheorem{example}{Example}}
\ifndef{\claim}{\newtheorem{claim}[theorem]{Claim}}

\newtheoremstyle{myclaims}   %
{\topsep}{\topsep}   %
{\slshape}   %
{}   %
{\bfseries\itshape}   %
{}   %
{ }   %
{\thmname{#1}\thmnumber{ \!#2}.}   %
{\theoremstyle{myclaims}

}

\newtheoremstyle{anystatement}{\topsep}{\topsep}{\itshape}{}{\bfseries}{.}{ }{\anystatementname}
{\theoremstyle{anystatement}}

\newcommand{\anystatementname}{}

\newcounter{tmp_id_cnt}
\newcommand{\AuxNew}[4][]{#2{#3}[1][*]            %
{\ifthenelse{\equal{*}{##1}}   %
{\Ensuremath{#1{#4}}}       %
{\ifthenelse{\equal{b}{##1}}   %
{\Ensuremath{\mathbf{#4}}}  %
{\ifthenelse{\equal{}{##1}}    %
{\IfMathMode{#1{#4}}{#4}}{}}}}}

\newcommand{\newident}[3][*]{\ifthenelse{\equal{*}{#1}}  %
{\AuxNew[\mathit]{\newcommand}{#2}{#3}}   %
{\mydef{#2}{\Ensuremath{\mathit{#3}}}}}   %

\newcommand{\newmat}[3][*]{\ifthenelse{\equal{*}{#1}}     %
{\AuxNew{\newcommand}{#2}{#3}}    %
{\mydef{#2}{\Ensuremath{#3}}}}       %

\newcommand{\providemat}[3][*]{\ifthenelse{\equal{*}{#1}}      %
{\AuxNew{\providecommand}{#2}{#3}}    %
{\mydef{#2}{\Ensuremath{#3}}}}           %

\newcommand{\providematarg}[2]{         %
\providecommand{#1}[1][]{\Ensuremath{#2}}}      

\newcommand{\newmatop}[2]{\mydef{#1}{\operatorname{#2}}}

\newcommand{\MyMakeTheoMacros}[3]{
\newcommand{#2}[2][]{\ifthenelse{\equal{}{##1}}
{\begin{#1} ##2 \end{#1}}
{\begin{#1}\label{##1} ##2\end{#1}}}
\newcommand{#3}[3][]{\ifthenelse{\equal{}{##1}}
{\begin{#1}{\e{##2}} ##3 \end{#1}}
{\begin{#1}{\e{##2}}\label{##1} ##3\end{#1}}}
}

\newcommand{\MyMakeDupTheoMacros}[8]{
\MyMakeTheoMacros{#1}{#2}{#3}
\newcommand{#4}[3]{
\newcommand{##2}{##3}
\begin{#1}\label{##1} ##2\end{#1}}
\newcommand{#5}[4]{
\newcommand{##2}{##4}
\begin{#1}{\e{##3}}\label{##1} ##2\end{#1}}
\newcommand{#8}[2]{\def\my_tmp_id{my_tmp_id_\arabic{tmp_id_cnt}}
\newtheorem*{\my_tmp_id}{#7~\ref{##1}}
\begin{\my_tmp_id} ##2 \end{\my_tmp_id}\stepcounter{tmp_id_cnt}}
\newcommand{#6}[6]{
#2[##1]{##2}

##3
\prf[#7~\ref{##1}]{##6} \newcommand{##5}{}

}
}

\newcommand{\MyMakeRefMacros}[3]{\newcommand{#1}[2][]
{\ifthenelse{\equal{}{##1}}{#2~\ref{##2}}{#3~\ref{##1} and~\ref{##2}}}}

\newcommand{\MyMakeEqRefMacros}[3]{\newcommand{#1}[2][]
{\ifthenelse{\equal{}{##1}}{#2~\eqref{##2}}{#3~\eqref{##1} and~\eqref{##2}}}}

\newcommand{\abstr}[1]{
\begin{abstract}
#1
\end{abstract}}

\newcommand{\bibentry}[8]{

\bibitem[\nospell{#8}]{#1} {\textup #3}. 
\ifthenelse{\equal{}{#6}}
{\newblock \textrm{#4.} \newblock {\em #5}, #7.}
{\newblock \textrm{#4.} \newblock {\em #5, #6}, #7.}
}

\MyMakeTheoMacros{fact}{\fct}{\nfct}

\MyMakeRefMacros{\fctref}{Fact}{Facts}

\MyMakeTheoMacros{observation}{\obs}{\nobs}

\MyMakeRefMacros{\obsref}{Observation}{Observations}

\MyMakeDupTheoMacros{lemma}
{\lem}{\nlem}{\lemdup}{\nlemdup}{\lemapp}{Lemma}{\lemrep}

\MyMakeRefMacros{\lemref}{Lemma}{Lemmas}

\newcommand{\fakelemref}[1]{Lemma~{#1}}

\MyMakeDupTheoMacros{corollary}
{\crl}{\ncrl}{\crldup}{\ncrldup}{\crlapp}{Corollary}{\crlrep}

\MyMakeRefMacros{\crlref}{Corollary}{Corollaries}

\MyMakeTheoMacros{proposition}{\prp}{\nprp}

\newtheorem*{prp*}{\e{Proposition}}

\MyMakeRefMacros{\prpref}{Proposition}{Propositions}

\MyMakeDupTheoMacros{my_claim}
{\clm}{\nclm}{\clmdup}{\nclmdup}{\clmapp}{Claim}{\clmrep}

\MyMakeRefMacros{\clmref}{Claim}{Claims}

\MyMakeDupTheoMacros{theorem}
{\theo}{\ntheo}{\theodup}{\ntheodup}{\theoapp}{Theorem}{\theorep}

\MyMakeRefMacros{\theoref}{Theorem}{Theorems}

\newcommand{\faketheoref}[1]{Theorem~{#1}}

\MyMakeTheoMacros{definition}{\defi}{\ndefi}

\MyMakeRefMacros{\defiref}{Definition}{Definitions}

\MyMakeTheoMacros{problem}{\prob}{\nprob}

\MyMakeRefMacros{\probref}{Problem}{Problems}

\MyMakeTheoMacros{protocol}{\prot}{\nprot}

\MyMakeRefMacros{\protref}{Protocol}{Protocols}

\providecommand{\qedsymbol}{\square}

\newcommand{\prf}[2][]{\ifthenelse{\equal{}{#1}}                %
{\begin{proof}\renewcommand{\qedsymbol}{$\blacksquare$}        %
#2 \end{proof}}                  %
{\begin{proof}[Proof of #1]      %
\renewcommand{\qedsymbol}{$\blacksquare_{\mbox{\it{\scriptsize{#1}}}}$}  %
#2 \end{proof}}
}

\newcommand{\sect}[2][]{\ifthenelse{\equal{}{#1}}
{\section{#2}}
{\section{#2}\label{#1}}}

\newcommand{\ssect}[2][]{\ifthenelse{\equal{}{#1}}
{\subsection{#2}}
{\subsection{#2}\label{#1}}}

\newcommand{\sssect}[2][]{\ifthenelse{\equal{}{#1}}
{\subsubsection{#2}}
{\subsubsection{#2}\label{#1}}}

\MyMakeRefMacros{\chref}{Chapter}{Chapters}

\MyMakeRefMacros{\sref}{Section}{Sections}

\MyMakeRefMacros{\ssref}{Subsection}{Subsections}

\MyMakeRefMacros{\sssref}{Subsection}{Subsections}

\definecolor{DarkGreen}{rgb}{0,0.45,0.08}

\newcommand{\IfMathMode}[2]{\ifmmode{#1}\else{#2}\fi}

\newcommand{\Ensuremath}{\ensuremath}

\newcommand{\fbr}[1]{\IfMathMode   %
{#1}{$#1$}}                              %

\newcommand{\fnbr}[1]{\mbox{\fbr{#1}}}   %

\newcommand{\fla}[2][*]{\ifthenelse{\equal{}{#1}}{\fbr{#2}}{\fnbr{#2}}}

\newcommand{\mat}[2][]{\ifthenelse{\equal{}{#1}}    %
{ \begin{displaymath} #2 \end{displaymath} }       %
{ \begin{equation} \label{#1} #2 \end{equation} } %
}

\newcommand{\matal}[2][]{\mat[#1]{\begin{aligned} #2 \end{aligned}}}

\newcommand{\f}{\fla}

\newcommand{\m}{\mat}
\newcommand{\mal}{\matal}

\newcommand{\mac}{\substack}    %

\newcommand{\twocase}[4]   %
{\begin{cases}#1 &\txt{#2}\\#3 &\txt{#4}\end{cases}}

\MyMakeEqRefMacros{\equref}{Equation}{Equations}

\MyMakeEqRefMacros{\expref}{Expression}{Expressions}

\MyMakeEqRefMacros{\inequref}{Inequality}{Inequalities}

\newcommand{\bracref}[1]{(\ref{#1})}

\newcommand{\bref}{\bracref}

\MyMakeRefMacros{\figref}{Figure}{Figures}

\providecommand{\middle}{\big}

\newcommand{\chs}{\genfrac(){0cm}{}}   %

\newmatop{\poly}{poly}
\newmatop{\sg}{sg}
\newmatop{\tr}{tr}

\newcommand{\h}[2][]{\ifthenelse{\equal{}{#2}}   %
{\mathop{\mathbf{H}}_{#1}}                      %
{\mathop{\mathbf{H}}_{#1}{\left[{#2}\right]}}}
\newcommand{\hh}[3][]{\mathop{\mathbf{H}}_{#1}   %
{\left[{#2}\middle|\vphantom{|_1^1}{#3}\right]}}

\newcommand{\KL}[2]{d_{KL}\llp{#1}\middle|\middle|{#2}\rrp}

\providecommand{\E}[2][]{\mathop{\mathbf{E}}_{#1}{\left[{#2}\right]}}
\newcommand{\Ee}[3][]{\mathop{\mathbf{E}}_{#1}{\left[{#2}\middle|\vphantom{|_1^1}{#3}\right]}}

\newcommand{\Var}[1]{\mathop{\mathbf{Var}}{\left[{#1}\right]}}

\newcommand{\PR}[2][]{\mathop{\mathbf{Pr}}_{#1}{\left[{#2}\right]}}

\newcommand{\U}[1][]{\ifthenelse{\equal{}{#1}} %
{{\cal U}}                                    %
{{\cal U}_{#1}}}
\newcommand{\Uu}[2][]{\ifthenelse{\equal{}{#1}}  %
{{\cal U}^{#2}}                               %
{{\cal U}_{#1}^{#2}}}

\newcommand{\pss}[1][]{\nospell{\ifthenelse{\equal{}{#1}} %
{\mbox{'s}}                                              %
{\fla{#1}\mbox{'s}}}}

\newcommand{\pl}[1][]{\nospell{\ifthenelse{\equal{}{#1}}  %
{\!\stackrel'{}\mbox{s}}                                 %
{\fla{#1\!\stackrel'{}}\mbox{s}}}}

\newcommand{\ord}[1][]{\nospell{\ifthenelse{\equal{}{#1}}
{\mbox{'th}}
{\ifthenelse{\equal{1}{#1}}{$1$\mbox{'st}}{\ifthenelse{\equal{2}{#1}}{$2$\mbox{'nd}}{\ifthenelse{\equal{3}{#1}}{$3$\mbox{'rd}}{\fla{#1}\mbox{'th}}}}}}}

\newmat[]{\llp}{\left(}
\newmat[]{\rrp}{\right)}
\providecommand{\lrp}[2][]{\ifthenelse{\equal{}{#1}}  %
{\Ensuremath{\left(#2\right)}}                       %
{\Ensuremath{\left(#2\middle)_{#1}\right.}}}

\newmat[]{\lla}{\left\langle}
\newmat[]{\rra}{\right\rangle}

\newmat[]{\llc}{\left\lceil}
\newmat[]{\rrc}{\right\rceil}

\newmat[]{\dt}{\cdot}
\newmat[]{\tm}{\cdot}
\newmat[]{\xor}{\oplus}
\newmat[]{\sbs}{\subset}
\newmat[]{\eps}{\varepsilon}
\newmat[]{\deq}{\stackrel{\textrm{def}}{=}}
\newmat[]{\ii}{\rm i}     %

\providemat{\QQ}{\mathbb{Q}}
\providematarg{\NN}{\ifthenelse{\equal{}{#1}}  %
{\mathbb{N}}                                  %
{\mathbb{N}_{#1}}}
\providematarg{\CC}{\ifthenelse{\equal{}{#1}}  %
{\mathbb{C}}                                  %
{\mathbb{C}^{#1}}}
\providematarg{\RR}{\ifthenelse{\equal{}{#1}}  %
{\mathbb{R}}                                  %
{\mathbb{R}^{#1}}}

\newcommand{\Den}[1]{\Ensuremath{\mathbf{Den}[#1]}}

\newmat[]{\ds}{\dots}
\newmat[]{\dc}{,\ds,}
\newmat[]{\dcirc}{\circ\ds\circ}

\newcommand{\itemi}[2][]{\ifthenelse{\equal{}{#1}}
{\begin{itemize} #2 \end{itemize}}
{\begin{itemize}[#1] #2 \end{itemize}}}

\newcommand{\wrt}{w.r.t.\ }	 %

\newcommand{\wlg}{w.l.g.\ }	 %

\newcommand{\ie}{i.e., }	 %

\newcommand{\eg}{e.g., }	 %

\newcommand{\fr}[3][*]{                       %
\ifthenelse{\equal{*}{#1}}                    %
{\frac{#2}{#3}}{}                           %
\ifthenelse{\equal{}{#1}}                     %
{\left.#2\middle/#3\right.}{}               %
\ifthenelse{\equal{b_}{#1}}                   %
{\left.\left(#2\right)\middle/#3\right.}{}  %
\ifthenelse{\equal{_b}{#1}}                   %
{\left.#2\middle/\left(#3\right)\right.}{}  %
\ifthenelse{\equal{bb}{#1}}                   %
{\left.\left(#2\right)\middle/\left(#3\right)\right.}{}
}

\newcommand{\set}[2][]{\ifthenelse{\equal{}{#1}}               %
{\Ensuremath{\left\{#2\right\}}}                             %
{\Ensuremath{\left\{#2\middle\bracevert #1\right\}}}}  %

\newcommand{\Min}[2][]{\ifthenelse{\equal{}{#1}}   %
{\Ensuremath{\min{\left\{#2\right\}}}}             %
{\Ensuremath{\min{\left\{#2\middle\bracevert #1\right\}}}}}  %

\newcommand{\Max}[2][]{\ifthenelse{\equal{}{#1}}   %
{\Ensuremath{\max{\left\{#2\right\}}}}             %
{\Ensuremath{\max{\left\{#2\middle\bracevert #1\right\}}}}}  %

\newcommand{\Maxx}[3][]{\ifthenelse{\equal{}{#1}}   %
{\Ensuremath{\max_{#2}{\left\{#3\right\}}}}             %
{\Ensuremath{\max_{#2}{\left\{#3\middle\bracevert #1\right\}}}}}  %

\newcommand{\newfunction}[2]{                          %
\newcommand{#1}[2][*]{\ifthenelse{\equal{*}{##1}}    %
{\Ensuremath{#2{\left(##2\right)}}}                   %
{#2(##2)}}                                          %
}                                                     %

\newfunction{\asO}{O}
\newfunction{\aso}{o}
\newfunction{\asOm}{\Omega}
\newfunction{\asom}{\omega}

\providecommand{\ket}[1]{\Ensuremath{\left|#1\rra}}
\providecommand{\bra}[1]{\Ensuremath{\lla #1\right|}}

\newcommand{\kbra}[2][]{\ifthenelse{\equal{}{#1}}                              %
{\Ensuremath{\left|#2\middle\rangle\hspace{-2.5pt}\middle\langle #2\right|}}  %
{\Ensuremath{\left|#1\middle\rangle\hspace{-2.5pt}\middle\langle #2\right|}}  %
}
\newcommand{\bket}[3][]{\ifthenelse{\equal{}{#1}} %
{\Ensuremath{\lla #2\middle|#3\rra}}
{\Ensuremath{\lla #2\middle|#1\middle|#3\rra}}
}

\providecommand{\ip}[2]{\Ensuremath{\lla #1,#2\rra}}

\mydef{\01}{\set{0,1}}
\mydef{\l(}{\llp}
\mydef{\r)}{\rrp}

\newcommand{\sz}[2][]{\ifthenelse{\equal{}{#1}}       %
{\Ensuremath{\left|#2\right|}}                       %
{\Ensuremath{\left|#2\middle|_{#1}\right.}}}
\providecommand{\norm}[2][]{\ifthenelse{\equal{}{#1}} %
{\Ensuremath{\left\|#2\right\|}}                     %
{\Ensuremath{\left\|#2\middle\|_{#1}\right.}}}

\providecommand{\ceil}[2][*]{\ifthenelse{\equal{}{#1}}  %
{\lceil #2 \rceil}                                     %
{\llc #2 \rrc}}

\newcommand{\fn}{\footnote}

\newcommand{\nin}{\not\in}   %

\newcommand{\e}{\emph}

\newcommand{\ul}[1]{\underline{#1}}  %

\newcommand{\txt}[1]{\textrm{#1}}   %

\newcommand{\tb}{\quad}
\newcommand{\tbbb}{\qquad\qquad}

\usepackage{hyphenat}   %

\date{}
\newmat{\calE}{\mathcal{E}}

\newmat[]{\Ub}{\U[\txt{bas}]}
\newmat[]{\Prho}{P_{V\sim\Ub}(\rho)}

\newident{\ECp}{\calE_{pure}^C}
\newident{\ECm}{\calE_{mix}^C}
\newident{\Ecla}{\calE_{cla}}

\title{Quantum Fingerprints that Keep Secrets}

\author{Dmitry Gavinsky%
\thanks{NEC Laboratories America, Inc., Princeton, NJ, U.S.A.}
\and Tsuyoshi Ito%
\thanks{Institute for Quantum Computing and David R.~Cheriton School of Computer Science, University of Waterloo, Waterloo, Ontario, Canada.}
}

\begin{document}

\maketitle

\thispagestyle{empty}

\abstr{We introduce a new type of cryptographic primitive that we call \e{hiding fingerprinting}.

A (quantum) fingerprinting scheme translates a binary string of length $n$ to $d$ (qu)bits, typically $d\ll n$, such that given any string $y$ and a fingerprint of $x$, one can decide with high accuracy whether $x=y$.
Classical fingerprinting schemes cannot hide information very well:\ a classical fingerprint of $x$ that guarantees error $\le\eps$ necessarily reveals \asOm{\log(1/\eps)} bits about $x$.
We call a scheme \e{hiding} if it reveals \aso{\log(1/\eps)} bits; accordingly, no classical scheme is hiding.

For any constant $c$, we construct two kinds of hiding fingerprinting schemes, both mapping $x\in\01^n$ to \asO{\log n} qubits and guaranteeing one-sided error probability at most $1/n^c$.
The first kind uses pure states and leaks at most \asO1 bits, and the second kind uses mixed states and leaks at most $1/n^c$ bits, where the ``leakage'' is bounded via accessible information.
The schemes are computationally efficient.

Our mixed-state scheme is optimal, as shown via a generic strategy that extracts $1/\poly(n)$ bits from any fingerprint over \asO{\log n} qubits.

Our results have a communication complexity interpretation.
We give quantum protocols for the equality problem in the models of \e{one-way communication} and \e{simultaneous message passing} that have communication cost~\asO{\log n} and offer \e{hiding guarantees} that cannot be matched by classical protocols of any cost.

Some of the technical lemmas in this work might be of independent interest.}

\setcounter{page}{1}

\sect{Introduction}

Cryptography probably is the area that benefits most from replacing classical computers by quantum ones.
In particular, the most restricting classical ``axiom'' of computational cryptography, the one it owes its name to, can be partially removed:
With quantum protocols it is no longer true that virtually \e{any interesting cryptographic protocol can be safe only if computational limitations of a potential intruder are assumed}.

The famous \e{quantum key distribution} protocol by Bennett and Brassard~\cite{BB84_Qua_Cr} is a good example where the assumption that ``an intruder is computationally limited'' has been replaced by the assumption that \e{quantum mechanics is valid in our physical universe}.
And if we accept quantum mechanics, it is highly desirable to find more examples of quantum crypto-protocols with unconditional security guarantees:
Besides pleasing those of us who prefer to keep their secrets for themselves, such examples might shed more light on the nature of differences between quantum and classical information.

Informally speaking, the possibility to use quantum mechanics in order to achieve unconditional cryptographic security comes from the fact that, in general, quantum states are not ``cloneable'' (cf.~\cite{WZ82_A_Si}).
Sometimes it can be very challenging to use this property alone (not making any computational assumptions) in order to build a cryptographic primitive; moreover, some very tempting goals are already known to be beyond the reach (cf.~\cite{M97_Un_Se}).
It is the quest of quantum cryptography to understand what crypto-goals can be achieved in a universe where the laws of quantum mechanics are valid.

\ssect[ss_claqua_hid]{Fingerprints and their hiding properties}

In this paper we will give a new example of a quantum crypto-primitive that is not achievable classically.
We call it \e{hiding fingerprints}.
Noticeably, hiding fingerprints are impossible classically even modulo arbitrarily strong consistent assumptions.

In the context of this work the meaning of (classical) \e{fingerprints} is as follows.
Given a binary string $x$ of length $n$, we want to (efficiently) produce its ``partial description'' by $d$ bits, typically with $d\ll n$, such that given only the description of $x$ and any $y\in\01^n$, one can test whether $x=y$ with high accuracy.
This can be achieved classically, for example by using a randomized mapping $x\to(s,h_s(x))$, where $h_s$ is chosen at random from a 2-universal family of hash functions ($s$ identifies $h_s$ inside the family).

\e{Quantum fingerprints} have been introduced by Buhrman, Cleve, Watrous and de~Wolf in~\cite{BCWW01}, however they were not treated as cryptographic primitives.
Generally speaking, an \e{\f n bits to \f d qubits quantum fingerprinting scheme} is a mapping from \f n-bit binary strings to density matrices in $2^d$-dimensional complex Hilbert space, such that when $\rho_x$ is the fingerprint of $x$ then given $\rho_x$ and $y$, one can decide with high confidence whether $x=y$.
Obviously, quantum fingerprints are a generalization of the classical ones.

Let \calE\ be a quantum fingerprinting scheme; we will be dealing with the following question.
Given $\rho_x$, how much classical information about $x$ can be ``extracted'' from it?
Formally, for any quantum measurement $P$, how large can be the mutual information between a random variable $X=x$ that is uniformly distributed over $\01^n$ and the outcome of $P$ applied to $\rho_x$?
The supremum of that value is called \e{the accessible information of \calE}.
In the special case when \calE\ is a classical scheme, its accessible information equals the mutual information between $X=x$ and a fingerprint of $x$ that \calE\ produces.

We will say that a fingerprinting scheme is \e{hiding} if its accessible information is \aso{\log(1/\eps)}.
This is the ``cryptographic ingredient'' that we add to the otherwise known notion of fingerprints.
\e{No classical fingerprinting scheme can be hiding}, as we see next.

Let \e{collision} be the event when a fingerprint of $x$ leads its holder to the conclusion that ``$x=y$'', even though the two strings are different.
Denote by $\eps_+$ the maximum collision probability, taken over all pairs $x\neq y$.
Let $\eps_-$ be the maximum, over all \pl[x], probability that the fingerprint holder declares ``$x\ne y$'', even though $y=x$.
Denote $\eps\deq\Max{\eps_+,\eps_-}$, this is the worst case error probability of the fingerprinting scheme.

Let \Ecla\ be a classical scheme that guarantees error at most $\eps$.
What happens when the holder of a fingerprint of $x$ loops through all $2^n$ possible values of $y$ and makes his best judgment whether $x=y$?
Let $A$ contain those \pl[y] where the guess was ``$x=y$'', then on the one hand, the expectation of $\sz A$ is at most $(2^n-1)\eps_++1$, and on the other hand, $x\in A$ with probability at least $1-\eps_-$.
Therefore, at least $(1-\eps_-)\log_2(1/\eps_+)\in\asOm{\log(1/\eps)}$ bits are leaked about $x$ by its fingerprint in \Ecla\ (unless $\eps=0$, in which case $n$ bits are leaked).
Accordingly, \Ecla\ is not hiding.

The same reasoning does not apply to the case of \e{quantum fingerprinting schemes}, where a binary string $x\in\01^n$ is mapped to a quantum state $\rho_x$, such that given any $y\in\01^n$ one can measure $\rho_x$, in order to decide with high accuracy whether $x=y$.
The argument fails because to make a guess whether $x=y$ one may be required to perform a quantum measurement, and such measurements can, in general, change the state of a quantum fingerprint in an irreversible way.
Alternatively, one can say that the ``looping trick'' cannot be used because $\rho_x$ is not necessarily cloneable.

From the practical point of view, hiding fingerprints shall be used when there is a need for a ``semi-trusted'' agent to be able to perform string recognition, but not to share with others the ability to recognize the target.
Putting it differently, hiding fingerprints allow to issue an ``authorization'' to perform certain pattern recognition limited number of times.

\ssect{Our results}

We construct new quantum fingerprinting schemes that hide information about $x$ in a way that cannot be achieved classically.
For any constant $c$, we construct two different schemes, both mapping $x\in\01^n$ to \asO{\log n} qubits and guaranteeing error probability at most $1/n^c$ when $x\neq y$ and no error when $x=y$.
The first scheme uses pure states and guarantees leaking of at most \asO1 bits; the second scheme uses mixed states and guarantees leaking of at most $1/n^c$ bits.
As follows from the previous argument, these results introduce a new type of cryptographic primitives that cannot be achieved classically.

Our schemes are computationally efficient.
Constructions themselves are probabilistic:
A description of a scheme includes polynomial number of random bits, and using uniformly chosen bits results in a good construction with all but exponentially small probability.
This random string can be viewed as a part of the scheme's definition, in particular it does not have to be kept in secret (e.g., it may be standardized to define a globally used scheme).\fn
{This is conceptually different from the role of randomness in any (nontrivial) classical fingerprinting scheme that inevitably depends on the assumption that the input strings $x$ and $y$ are chosen independently from the random seed used to build a fingerprint of $x$.}

The ``hiding guarantees'' of our mixed-state schemes are optimal.\fn
{Our optimality argument can probably be tuned to show that our pure-state construction is also optimal.
We have not pursued that direction, since the mixed-state schemes are a natural generalization of the pure-state ones, and therefore the interest of showing optimality of a pure-state construction within its own class would be limited.}
To demonstrate that we construct a generic strategy for extracting information from arbitrary quantum fingerprints.
This ``no-go'' result remains valid for several weaker notions of fingerprinting schemes than what we construct (e.g., for schemes with two-sided error; see~\sref{s_opti} for more).

More formally, our main results are (cf.~\theoref[t_Main]{t_optimal}):

\theo{For any constant $c$ there exist quantum fingerprinting schemes that
\itemi{
\item map \f n-bit strings to mixed states of \asO{\log n} qubits and whose error probability and accessible information are both bounded by $1/n^c$;
\item map \f n-bit strings to pure states of \asO{\log n} qubits, whose error probability is bounded by $1/n^c$ and accessible information is \asO1.
}
The schemes are computationally efficient and have one-sided error with $\eps_-=0$ (answers ``$x\neq y$'' are always true).

Any quantum fingerprinting scheme that uses $d$ qubits and guarantees error below $1/2-\asOm1$ has accessible information $2^{-O(d)}$.
}

To the best of our knowledge, hiding fingerprints cannot be obtained via classical reduction to any previously known quantum cryptographic primitive.

Some of our technical contributions might be of independent interest.

\sssect{Communication complexity perspective}

The notion of quantum fingerprints has been introduced in~\cite{BCWW01} mainly in the context of \e{communication complexity}.
The main conceptual contribution of the present work is to view quantum fingerprints as a cryptographic primitive.
Nevertheless, our results can be interpreted in the language of communication complexity, as follows.

The most common communication complexity scenario is the one where two players, Alice and Bob, receive two parts of input, $x$ and $y$, respectively.
The players communicate in order to compute the value of certain function $f(x,y)$, trying to minimize the amount of communication.
Various models exist that define the constraints that Alice and Bob have to obey when they compute $f(x,y)$.
Relevant to us are the following two:\itemi{
\item \e{One-way communication} is a model where Alice sends a single message to Bob, who has to give an answer based on that message and his input $y$.
\item \e{Simultaneous Message Passing (SMP)} is a model involving a third participant, \e{a referee}.
Here both Alice and Bob send one message each to the referee, who has to give an answer based on the received messages.\fn
{We consider the version of SMP without shared randomness.}
}
In both the cases the players are computationally unlimited, and the \e{cost} of a communication protocol equals the total number of sent bits.
Quantum analogues of the models can be defined, where players send qubits and locally perform arbitrary unitary transformations.

One of the most basic communication problems corresponds to the \e{equality} predicate, where the goal of the players is to decide whether $x=y$.
In general, fingerprinting schemes can be naturally viewed as solutions to the equality problem, as follows.\fn
{This was used in~\cite{BCWW01} to demonstrate exponential separation between the quantum and the classical versions of the SMP model.}

\ul{In the model of SMP}, Alice and Bob both send the fingerprints of, respectively, $x$ and $y$ to the referee.
Then the referee performs the swap test that would always return ``equal'' if $x=y$ and would have positive constant probability of returning ``not equal'' if $x\neq y$.
Thus, he can answer whether $x=y$ with one-sided constant error.

If such a protocol is based on our pure-state hiding fingerprinting scheme then its cost is \asO{\log n}.
It follows from the hiding guarantees of our schemes that this protocol is also \e{hiding}:\ an ``eavesdropper'' can learn \e{at most \asO{1} bits} of information about the input $(x,y)$.

On the other hand, as shown by Newman and Szegedy~\cite{NS96_Pub}, the classical SMP-complexity of checking equality with constant error probability is~\asOm{\sqrt n}.
Their argument readily implies that any classical protocol leaks \e{at least \asOm{\sqrt n} bits} about the input.
Moreover, this holds for classical protocols of any cost!

\ul{In the model of one-way communication}, our mixed-state hiding fingerprinting scheme translates trivially to a protocol of cost \asO{\log n} that solves the equality problem with error at most $1/\poly$ and leaks at most $1/\poly$ bits about the input.
On the other hand, our classical impossibility argument implies that any classical protocol that solves the equality problem with error $\eps$ necessarily leaks \asOm{\log(1/\eps)} bits about the input, and this is true for protocols of any cost.

\sect{Preliminaries and more}

Here we state only those technical lemmas that are relevant for the first part of the paper (construction and analysis of the new fingerprinting schemes).
Lemmas that will be used only in the second part of the paper (showing optimality of our schemes) will be stated is \sref{ss_opti_tech}. 

We write $\exp(x)$ and $\sg(x)$ to denote $e^x$ and  $(-1)^x$, respectively.
We write $\log$ to denote the natural logarithm and $\log_2$ for the logarithm to the base $2$.
We denote $\ii=\sqrt{-1}$ (to be distinguished from the variable $i$).

We let $\NN=\set{1,2\dc}$ and $[i]=\set{1,2\dc i}$.
We often implicitly assume the natural correspondence between the elements of $[2^n]$ and those of $\01^n$.
For any finite set $A$ we let $\U[A]$ denote the uniform probability distribution over the elements of $A$.

We use $\circ$ to denote concatenation of strings.
For any set $A$ and $x\in A^n$ we will write $x_i$ to address the \ord[i] position of $x$; more generally, $x_{i_1\dc i_k}\deq x_{i_1}\dcirc x_{i_k}$ for $(i_1\dc i_k)\in[n]^k$.
For two strings $x$ and $y$ of the same length, we will let $d_H(x, y)\deq\sz{\set[x_i\neq y_i]{i}}$ stand for the Hamming distance.

For $D\in\NN$, we write $I_D$ to denote the identity operator over $\CC^D$.
For a $D\times D$ matrix~$X$, we denote the trace norm of~$X$ by~$\norm[1]{X}=\tr\l(\sqrt{X^*X}\r)$, and the operator norm of~$X$ by~$\norm{X}=\Max[\sz v=1]{\sz{Xv}}$.

We will mostly use Dirac's ``bra-ket'' notation for pure quantum states, but sometimes we will find it convenient to switch to the standard notation (e.g., both $v$ and $\ket v$ will be used to denote the same unit vector in a Hilbert space).
We will be addressing mixed states via their density matrices, and denote by \Den D the subset of $\CC[D\times D]$ corresponding to density matrices.

\ssect{Random variables and their concentration}

The Hoeffding bound will be one of our basic tools, we will use it in the following form (\faketheoref{2.5} in \cite{D98_Con}):

\nlem[l_Hoef]{(Hoeffding bound)}{Let the random variables $X_1\dc X_n$ be mutually independent, satisfying $\E X_i=\mu_i$ and $a_i\le X_i\le b_i$ for some constants $a_i$ and $b_i$ for all $i$.
Then for any $t>0$,
\m{\PR{\sz{\sum X_i-\sum\mu_i}\ge t}
\le2\exp\l(\fr{-2t^2}{\sum(b_i-a_i)^2}\r).}
}

The following lemma can be viewed as a generalization of the Hoeffding bound to the case of random variables taking values in \CC.\fn
{We view \CC\ as a vector space isometric to \RR[2].
For the general case of random variables taking values in an Euclidean space there are known ``dimension-independent'' bounds.
We do not use one of those, instead we state \lemref{l_Hoef_c} whose proof is ``dimension-dependent'' but the final expression is more convenient for our purposes.}

\lem[l_Hoef_c]{Let the random variables $X_1\dc X_n$ take values in \CC\ and be mutually independent, satisfying $\E X_i=0$ and $\sz{X_i}\le c_i$ for some constants $c_i$ for all $i$.
Then for any $t>0$,
\m{\PR{\sz{\sum X_i}\ge t}
\le4\exp\l(\fr{-t^2}{4\sum\sz{c_i}^2}\r).}
}

\prf{By the Hoeffding bound (\lemref{l_Hoef}), for any $u>0$
\m{\PR{\Re\l(\sum X_i\r)\ge u},~
\PR{\Im\l(\sum X_i\r)\ge u}
\le2\exp\l(\fr{-u^2}{2\sum\sz{c_i}^2}\r).}
As $|\sum X_i|\ge t$ implies that either $\Re(\sum X_i)\ge\sqrt{t^2/2}$ or $\Im(\sum X_i)\ge\sqrt{t^2/2}$, the result follows.}

The next statement will be very convenient for proving upper bounds on expected values of random variables.

\lem[l_E_subst]{Let $f$ be a monotone non-decreasing function taking non-negative values, and let $Y$ and $\tilde Y$ be random variables satisfying
\f{\PR{\tilde Y\ge y}\ge\PR{Y\ge y}}
for every $y$ such that $f(y)>0$.
If $\E{f(\tilde Y)}<\infty$ then
\f{\E{f(\tilde Y)}\ge\E{f(Y)}.}
}

\prf{Let $Z\deq f(Y)$ and $\tilde Z\deq f(\tilde Y)$.
Then $Z\ge0$ and for every $z\ge0$ it holds that
\m{\PR{\tilde Z\ge z}\ge\PR{Z\ge z}.}
Therefore,
\m{\E{Z}=  \int_0^\infty \PR{Z\ge z}dz\le\int_0^\infty \PR{\tilde{Z}\ge z}dz = \E{\tilde{Z}},}
as required.
}

Our next goal is to prove yet another generalization of the Hoeffding bound.
We will use a modification of the standard method for proving such bounds, namely the ``Bernstein's trick''.
The next lemma is the main technical ingredient for that.

\lem[l_E_exp]{Let $Y$ be a random variable satisfying $\E Y=0$, $Y\ge a$ and $\PR{Y\ge y}\le\beta\exp(-\alpha(y-a))$ for all $y\ge a$ and some constants $a\le0$, $\beta\ge1$ and $\alpha>0$.
Then for every $h\in(0,\alpha/2]$ and $c\in(0,2]$,
\m{\E{\exp(hY)}
\le c+\exp\llp\fr{\llp\log\fr{2\beta}c\rrp^2}{2\alpha^2}h^2\rrp
\le\exp\llp c+\fr{\llp\log\fr{2\beta}c\rrp^2}{2\alpha^2}h^2\rrp.}
}

\prf{Denote by $E_b$ the event that $\lla Y\le b\rra$, were $b\ge a+\fr{\log\beta}{\alpha}$ is a constant, and let $I_b$ be the Boolean indicator of $E_b$.
Then
\m[m_exp_1]{\E{\exp(hY)}=\E{I_b\dt\exp(hY)}+\E{(1-I_b)\dt\exp(hY)}.}

Let $Y_1$ be a random variable distributed as $Y$ modulo $E_b$.
Then $\E{I_b\dt\exp(hY)}\le\E{\exp(hY_1)}$, $\E{Y_1}\le\E Y=0$, and $a\le Y_1\le b$.
A standard result from the theory of concentration bounds (\eg see \fakelemref{2.6} in \cite{D98_Con}) implies that
\m{\E{\exp(hY_1)}\le\exp\llp\fr{(b-a)^2}8h^2\rrp.}

Let $Y_2$ be a random variable satisfying $\PR{Y_2\ge y}=\beta\exp(-\alpha(y-a))$ for all $y\ge b$.
Then \lemref{l_E_subst} implies that
\mal{\E{(1-I_b)\dt\exp(hY)}
&\le\E{(1-I_b)\exp(hY_2)}
=\int_b^{\infty}\exp(hy)\dt\beta\alpha\exp(-\alpha(y-a))\,dy\\
&\le\beta\alpha\int_b^{\infty}\exp((h-\alpha)(y-a))\,dy
\le\beta\alpha\int_b^{\infty}\exp(-\fr{\alpha}2(y-a))\,dy.}

From \bref{m_exp_1},
\mal{\E{\exp(hY)}
&\le\exp\llp\fr{(b-a)^2}8h^2\rrp
+\beta\alpha\int_b^{\infty}\exp(-\fr{\alpha}2(y-a))\,dy\\
&=\exp\llp\fr{(b-a)^2}8h^2\rrp+2\beta\exp(-\fr{\alpha}2(b-a)).}
This holds for every $b\ge a+\fr{\log\beta}{\alpha}$, therefore
\m{\E{\exp(hY)}
\le\Min[b'\ge\fr{\log\beta}{\alpha}]
{\exp\llp\fr{b'^2}8h^2\rrp+2\beta\exp(-\fr{\alpha}2b'))}.}

Let $c\in(0,2]$ be any, and choose $b'=\fr2{\alpha}\log\fr{2\beta}c$.
Then $2\beta\exp(-\fr{\alpha}2b')=c$ and 
\m{\E{\exp(hY)}
\le\exp\llp\fr{b'^2}8h^2\rrp+c
=\exp\llp\fr{\llp\log\fr{2\beta}c\rrp^2}{2\alpha^2}h^2\rrp+c,}
which is the first inequality stated in the lemma.
Finally,
\m{c+\exp\llp\fr{\llp\log\fr{2\beta}c\rrp^2}{2\alpha^2}h^2\rrp
\le(1+c)\exp\llp\fr{\llp\log\fr{2\beta}c\rrp^2}{2\alpha^2}h^2\rrp
\le\exp\llp c+\fr{\llp\log\fr{2\beta}c\rrp^2}{2\alpha^2}h^2\rrp,}
as $\log(1+c)<c$ for $c>0$.
}

We are ready to prove a new concentration bound, that can be viewed as a ``less demanding'' analogue of the Hoeffding bound.

\theo[t_Cher+]{Let the random variables $X_1\dc X_n$ be mutually independent, satisfying $\E X_i=\mu$, $X_i\ge a$ and $\PR{X_i\ge x}\le\beta\exp(-\alpha(x-a))$ for all $x\ge a$, $i\in [n]$ and some constants $a\le0$, $\alpha>0$ and $\beta\ge1$.
Let $S_n\deq\sum X_i$ for $i\in [n]$.
Then for every $t\in(0,\fr1{7\alpha}]$,
\m{\PR{\fr1nS_n\ge\mu+t}\le
\exp\llp-\fr{nt^2\alpha^2}{244\llp\log\fr{\beta}{t\alpha}\rrp^2}\rrp.}
}

\prf{By \lemref{l_E_exp}, for any $h\in(0,\alpha/2]$ and $c\in(0,2]$
\m{\E{\exp\llp h(S_n-n\mu)\rrp}
=\prod\E{\exp\llp h(X_i-\mu)\rrp}
\le\exp\llp nc+\fr{n\llp\log\fr{2\beta}c\rrp^2}{2\alpha^2}h^2\rrp.}
By Markov's inequality,
\m{\PR{S_n\ge n\mu+nt}\le\exp(-hnt)\E{\exp\llp h(S_n-n\mu)\rrp}
\le\exp\llp nc+\fr{n\llp\log\fr{2\beta}c\rrp^2}{2\alpha^2}h^2-hnt\rrp.}

Let
\m{c_0\deq\fr{t^2\alpha^2}{122\llp\log\fr{\beta}{t\alpha}\rrp^2}~\txt{and}~
h_0\deq\fr{t\alpha^2}{\llp\log\fr{2\beta}{c_0}\rrp^2}.}
From $t\alpha\le\fr17$, it holds that $0<c_0<1$ and $0<h_0<\alpha/2$.
Thus we may substitute $h=h_0$ and $c=c_0$, still satisfying the requirements of \lemref{l_E_exp}.
So,
\m{\PR{\fr1nS_n\ge\mu+t}
\le\exp\llp nc_0-\fr{nt^2\alpha^2}{2\llp\log\fr{2\beta}{c_0}\rrp^2}\rrp.}
It can be seen\fn
{Let $x\deq t\alpha$ and $f(x,\beta)\deq\fr[]{c_0}{\fr{t^2\alpha^2}{4\llp\log\fr{2\beta}{c_0}\rrp^2}}$, then modulo $x\in(0,\fr17]$ and $\beta\ge1$ it is always true that $\fr{df}{d\beta}<0$.
Let $f'(x)\deq f(x,1)$, then $\fr{df'}{dx}>0$ and therefore $f(x,\beta)\le f(\fr17,1)<1$.}
that $t\alpha\le\fr17$ and $\beta\ge1$ imply $c_0<\fr[]{t^2\alpha^2}{4\llp\log\fr{2\beta}{c_0}\rrp^2}$, and therefore
\m{\PR{\fr1nS_n\ge\mu+t}\le\exp\l(-\fr{nc_0}2\r)
=\exp\l(-\fr{nt^2\alpha^2}{244\l(\log\fr{\beta}{t\alpha}\r)^2}\r),}
as required.}

\ssect{$\eps$-nets for pure states}

In our proof we will need a ``continuous analogue'' of the union bound:
Namely, for every $D\in\NN$ we want to have some sufficiently large $T$, such that if certain event $E(v)$ holds with probability at most $\delta$ for any fixed vector $v\in\CC^D$, then with probability at least $1-T\delta$ there is no $v'\in\CC^D$ such that $E(v')$ holds.
Of course, in general that is not possible for infinite domains like $\CC^D$; however, the situation can be helped if there exists a ``relaxed'' version of $E$, that we denote by $E'$, such that if $E(v)$ holds and $d(v,w)\le\eps$, where $d(\dt,\dt)$ is a measure of distance between vectors in $\CC^D$ and $\eps$ is sufficiently small, then $E'(w)$ must also hold.

Fix $\eps$ and let $W_{\eps}=\set{w_1\dc w_T}$ be a finite set of vectors from $\CC^D$, such that for every $v\in\CC^D$ there exists some $w_i\in W_{\eps}$ satisfying $d(v,w_i)\le\eps$ (such sets are commonly called \e{$\eps$-nets}).
Assume that for any fixed $v\in\CC^D$ the probability that $E'(v)$ holds is at most $\delta$.
Then, by the union bound, the probability that $E'(w)$ holds for some $w\in W_{\eps}$ is at most $T\delta$.
Now, if $E(v)$ holds for some $v\in\CC^D$, then $E'(w)$ holds for at least one $w\in W_{\eps}$, as the set contains an element at distance at most $\eps$ from $v$.
Therefore, the probability that $E(v)$ holds for some $v\in\CC^D$ is at most $T\delta$.

The notion of distance between vectors can be formalized in many different ways, depending on the nature of $E$ and $E'$.
The following definition serves our future goals.

\defi[d_eps-net]{For $\eps>0$, we call a set~$M\subseteq\CC^D$ of unit vectors an \emph{$\eps$-net for the set of pure states in~$\CC^D$ with respect to the trace distance}, if for every unit vector~$\ket{u}\in\CC^D$ there exists $\ket{v}\in M$, such that $\norm{\kbra{u}-\kbra{v}}_1 \le \eps$.}

The following lemma is a slight improvement over \fakelemref{II.4} of \cite{HLSW04_Ran_Qua} and \fakelemref{4} of \cite{BHLSW05_Rem_Pre}, where the upper bound on the size of the $\eps$-net was $(5/\eps)^{2D}$.

\lemdup{l_eps-net}{\lemEpsNet}{For every $0<\varepsilon\le2$,
there exists an $\varepsilon$-net for the set of pure states in~$\CC^D$ with respect to the trace distance whose size is at most $(4/\varepsilon)^{2(D-1)}$.}

The proof of the lemma is given in the appendix.

\sect[s_schemes]{New quantum fingerprinting schemes and their properties}

We will use the standard way to construct a (pure-state) quantum fingerprinting scheme based on a classical error-correcting code.
Namely, given a code $C$ from $n$ to $2^d$ bits, we will define, for every $a\in\01^n$, its fingerprint on $d$ qubits via $\ket{u_a}=\fr1{2^{d/2}}\sum_{i\in[{2^d}]}\sg(b_i)\ket{i}$, where $b=(b_1\dc b_{2^d})=C(a)$.

It would be very convenient for us to use a perfectly random code $C$; however we cannot afford that as we want our construction to be computationally efficient.
On the other hand, we can get an efficient construction by using a random linear $C$, however it turns out that such code would not be ``random enough'' for our needs (we need more randomness to guarantee that a scheme is hiding).\fn
{Note that in the context of quantum fingerprinting there is no need to ever decode the underlying classical code, that is why using a random linear code would be computationally feasible, despite the fact that no efficient decoding is known to exist for such codes.}
So, we define a new type of classical codes that still admit efficient encoding but contain more randomness than random linear codes.

\ssect[ss_qlc]{Random quasi-linear codes}

In the following definition we use $2^d$ to denote the codewords' length in order to make the notation more consistent throughout the paper.

\defi{Let $r,n,d\in\NN$, $r<n<2^d$.
An \e{$(n,r,{2^d})$-quasi-linear code} $C$ is represented by an ${2^d}$-tuple of $(n-r)$-bit vectors $(c_1\dc c_{2^d})$ and a $2^r$-tuple of ${2^d}$-bit vectors $(d_1\dc d_{2^r})$.
For every $a\in\01^n$ we denote $a^{(1)}\deq a|_{1\dc r}$, $a^{(2)}\deq a|_{r+1\dc n}$, and define  
\m{C(a)\deq d_{a^{(1)}}\xor\llp\ip{c_i}{a^{(2)}}\rrp_{i=1}^{2^d},}
where $\xor$ denotes bit-wise \e{xor}.}

In other words, $(d_1\dc d_{2^r})$ is an arbitrary code applied to the first $r$ bits of $a$ and $(c_1\dc c_{2^d})$ defines a linear code that is applied to the last $(n-r)$ bits; the actual encoding of $a$ is the xor of the two codewords.

For the rest of the paper we will write $x^{(1)}$ and $x^{(2)}$ to address, respectively, $x|_{1\dc r}$ and $x|_{r+1\dc n}$, when $n$, $r$ and $x\in\01^n$ are clear from the context.

Obviously, $C(a)$ can be computed efficiently when $r\in\asO{\log n}$ and $d\in\asO{\log(n)}$.
We call a quasi-linear code (uniformly) random if both $(c_1\dc c_{2^d})$ and $(d_1\dc d_{2^r})$ are selected uniformly at random.
We will denote this distribution by $\U[C]$ and write $C\sim\U[C]$ to say that $C$ is chosen uniformly at random (the values of the parameters $n$, $r$ and $d$ will be clear from the context).
Note that efficient description of such code is possible as long as $r\in\asO{\log n}$ and $d\in\asO{\log(n)}$.

The following property of random quasi-linear codes can be viewed as a generalization of the notion of minimal distance.
Denote $\gamma_C\deq\Max[a_1\neq a_2]{\sz{d_H\l(C(a_1),C(a_2)\r)-2^{d-1}}}$.
Then

\lem[l_min_gen]{For every $t>0$, $\PR[{C\sim\U[C]}]{\gamma_C\ge t}<2\exp\l(n+r-\fr{2t^2}{{2^d}}\r)$.}

\prf{Define $A_C\deq\set[a_1\ne a_2]{C(a_1)\xor C(a_2)}$.
Observe that $A_C=B_1\xor B_2\cup B_1\cup B_2$, where $\xor$ is element-wise, $B_1=\set[a_1,a_2\in\01^r; a_1\neq a_2]{d_{a_1}\xor d_{a_2}}$ and
\m{B_2
=\set[a_1,a_2\in\01^{n-r};a_1\neq a_2]
{\llp\ip{c_i}{a_1\xor a_2}\rrp_{i=1}^{n-r}}
=\set[0\neq a\in\01^{n-r}]{\llp\ip{c_i}a\rrp_{i=1}^{n-r}}.}
Direct counting reveals that $\sz{A_C}\le2^{n+r}$.

It is easy to see that for every $a_1\ne a_2$ the string $C(a_1)\xor C(a_2)$ is chosen uniformly at random from $\01^{2^d}$ when $C\sim\U[C]$.
By the Hoeffding bound (\lemref{l_Hoef}), for every $t>0$
\m{\PR[{C\sim\U[C]}]{\sz{d_H\l(C(a_1),C(a_2)\r)-2^{d-1}}\ge t}
\le2\exp\l(\fr{-2t^2}{{2^d}}\r),}
and the union bound implies the statement of the lemma.}

\ssect[s_pure]{Pure-state scheme}

For the rest of the paper we assume that $d\in\asO{\log n}$ and that $r\in\asO{\log n}$.

First, we define and analyze our fingerprinting scheme that uses pure states.
Afterwords (\sref{s_mixed}) we will consider a mixed-state scheme that can be viewed as a generalization.

\defi[d_ECp]{
Let $C$ be an $(n,r,{2^d})$-quasi-linear code, we denote by $\ECp$ the following fingerprinting scheme.
Every $a\in\01^n$ is mapped to 
\m{\ket{u_a}=\fr1{2^{d/2}}\sum_{i\in[2^d]}\sg(b_i)\ket{i},}
where $b=(b_1\dc b_{2^d})=C(a)$.
We call $\ket{u_a}$ the \e{fingerprint} of $a$.

Given $\ket{u_{a_1}}$ and any $a_2\in\01^n$, in order to check whether $a_1=a_2$ one should measure $\ket{u_{a_1}}$ \wrt the projective measurement \set{P_{a_2},I_{2^d}-P_{a_2}}, where $P_{a_2}=\kbra{u_{a_2}}$.
If the outcome corresponds to $P_{a_2}$ then ``$a_1=a_2$'' shall be returned, otherwise the guess should be ``$a_1\ne a_2$''.}

Note that the transformation $a\to\ket{u_a}$ can be computed efficiently as long as $C(a)$ is easy to compute for every $a$, and that the required projective measurement can be performed efficiently because $d\in\asO{\log(n)}$ and $\ket{u_{a_2}}$ is known.

Intuitively, the fingerprints corresponding to different pre-images should be nearly orthogonal.
This is formalized by the following lemma.

\lem[l_max_ip]{For $\set[a\in\01^n]{\ket{u_a}}$ defined over a randomly chosen $(n,r,{2^d})$-quasi-linear code $C$, for any $\delta>0$ it holds that $\Max[a_1\ne a_2]{\sz{\bket{u_{a_1}}{u_{a_2}}}}<\delta$ with probability at least $1-2\exp(n+r-\delta^22^{d-1})$.}

\prf{
\m{\sz{\bket{u_{a_1}}{u_{a_2}}}
=\fr1{2^d}\sz{\sum_{i\in[2^d]}\sg\l({b_1}_i+{b_2}_i\r)}
=\sz{2\fr{d_H\l(b_1,b_2\r)}{2^d}-1}\le
\fr{2\gamma_C}{2^d},}
where $b_1=C(a_1)$ and $b_2=C(a_2)$.
By \lemref{l_min_gen},
\m{\PR[{C\sim\U[C]}]{\fr{2\gamma_C}{2^d}\ge\delta}
<2\exp\l(n+r-\delta^22^{d-1}\r),}
as required.}

Now let us see that $\ECp$ is likely to be a valid fingerprinting scheme.

\lem[l_ECp_fp]{For $\ECp$ defined over a randomly chosen $(n,r,{2^d})$-quasi-linear code $C$, it holds that $\eps_-=0$ always and that $\eps_+<\delta$ with probability at least $1-2\exp\l(n+r-2^{d-1}\delta\r)$, for any $\delta>0$.}

\prf{Clearly, when $a_1=a_2$ the answer is always correct, that is $\eps_-=0$.
When, on the other hand, $a_1\ne a_2$ the probability of the wrong answer is $\sz{\bket{u_{a_1}}{u_{a_2}}}^2$, and therefore by \lemref{l_max_ip}, $\eps_+<\delta$ with probability at least $1-2\exp(n+r-2^{d-1}\delta)$, as required.}

Our next goal is to show that $\ECp$ defined over a randomly chosen quasi-linear code $C$ is hiding with high probability.
This will be done in stages.

Let us denote for every $a\in\01^n$:\ $\rho_a\deq\kbra{u_a}$, $\rho_a'\deq2^{d-n}\rho_a$, and for arbitrary $v\in\CC^{2^d}$, $\mu_v(a)\deq\bra v\rho_a'\ket v$.

We will see later (\lemref{l_avgstate}) that for almost all choices of~$C$ we have $\sum_a\rho_a'=I_{2^d}$, and therefore $\mu_{v}(a)$ is a probability distribution over $a\in\01^n$ for every fixed unit vector $v$.
Intuitively, this distribution corresponds to the ``view about $a$'' of a holder of $\rho_a$ who has measured it and got the outcome $\kbra{v}$.
Therefore, if originally $a$ was chosen uniformly then some sort of distance between $\mu_{v}$ and $\U[\01^n]$ should tell us how much has been learnt about $a$ as a result of the measurement.

The following technical statement is the key part of our upper bound on the accessible information for $\ECp$.

\lem[l_ECp_Eh]{Let $v\in\CC^{2^d}$ be a unit vector and $a_0\in\01^n$ be fixed, and assume that $\ECp$ is defined over an $(n,r,{2^d})$-quasi-linear code $C$, then
\m{\E[{C\sim\U[C]}]
{\Max{0,\mu_{v}(a_0)\log\l(2^n\mu_{v}(a_0)\r)}}<\fr{23}{2^n}.}
}

In the view of the intuition expressed above, it shouldn't be surprising that we want to prove this kind of statement.
Indeed, if $\mu_{v}$ is a probability distribution then $\sum_a\mu_{v}(a)\log\l(2^n\mu_{v}(a_0)\r)$ is the relative entropy between $\mu_{v}$ and $\U[\01^n]$.

\prf{Let
\m{\omega_v^a\deq\sz{\sum_{i\in[2^d]}\sg\l(\ip{a^{(2)}}{c_i}\xor {d_{a^{(1)}}}_i\r)v_i}^2,} 
then $\mu_{v}(a_0)=\fr{\omega_{v}^{a_0}}{2^n}$ and for every $t\ge0$,
\m[m_P_p]{\PR[C]{\mu_{v}(a_0)\ge\fr t{2^n}}
=\PR{\omega_{v}^{a_0}\ge t}
=\PR[{\beta_1\dc\beta_{2^d}\sim\U[\{-1,1\}]}]
{\sz{\sum\beta_i{v}_i}\ge\sqrt t}
\le4\exp\l(\fr{-t}4\r),}
where the inequality follows from \lemref{l_Hoef_c} and the fact that $\norm{v}=1$.

Define $g(x)\deq\Max{0,x\log(x)}$ and let $\tilde\mu$ be a random variable whose distribution satisfies $\PR{\tilde\mu\ge t}=4\exp(-t/4)\deq f(t)$ for $t\ge8\log2$.
Then
\m{\E{\Max{0,\mu_{v}(a_0)\log\l({2^n}\mu_{v}(a_0)\r)}}
\le\fr1{2^n}\E{g({2^n}\mu_{v}(a_0))}
\le\fr1{2^n}\E{g(\tilde\mu)},}
where the first inequality follows from the definition of $g(\dt)$ and the second one is by \lemref{l_E_subst} (whose requirements are implied by \bref{m_P_p} and \pss[g] definition).

Finally,
\m{\E{g(\tilde\mu)}
=\int_{8\log2}^{\infty}x\log(x)\l(-\fr{df}{dx}\r)dx
=\int_{8\log2}^{\infty}\exp(\log x+\log\log x-x/4)\,dx
<23,}
as required.}

At this point we suspend our analysis of $\ECp$ and turn to a mixed-state scheme $\ECm$.
Analysis of $\ECp$ will be resumed and merged with that of $\ECm$ in \sref{s_fa}.

\ssect[s_mixed]{Mixed-state scheme}

To define our mixed-state scheme we introduce another parameter $k\in\NN\cup\set0$, such that $2^k$ is the rank of every fingerprint (\ie $k=0$ corresponds to a pure-state scheme).
It will always be assumed, often implicitly, that $d\ge k$ and $r\ge k$ (the second assumption is probably less obvious, we need it for technical reasons). 

\defi[d_ECm]{
Let $C$ be an $(n+k,r,{2^d})$-quasi-linear code, where $d\ge k$ and $r\ge k$.
We denote by $\ECm$ the following fingerprinting scheme.
For every $x\in\01^{n+k}$ we let
\m{\ket{u_x}=\fr1{2^{d/2}}\sum_{i\in[2^d]}\sg(b_i)\ket{i},}
where $b=(b_1\dc b_{2^d})=C(x)$.
Every $a\in\01^n$ is mapped to
\m{\rho_a=\fr1{2^k}\sum_{i\in\01^k}\kbra{u_{i\circ a}}.}
We call $\rho_a$ the \e{fingerprint} of $a$.

Given $\rho_{a_1}$ and any $a_2\in\01^n$, in order to check whether $a_1=a_2$ one should measure $\rho_{a_1}$ \wrt the POVM measurement \set{P_{a_2},I_{2^d}-P_{a_2}}, where $P_{a_2}$ is the projection to the subspace of \RR[2^d] that is spanned by \set[i\in\01^k]{u_{i\circ a_2}}.
If the outcome corresponds to $P_{a_2}$ then ``$a_1=a_2$'' shall be returned, otherwise the guess should be ``$a_1\ne a_2$''.}

Note that when $k=0$ the above definition gives $\ECp$, and the notions of $\ket{u_a}$ and $\rho_a$ coincide with those considered in \sref{s_pure}.
To construct $\rho_a$, the holder of $a$ tosses $i\sim\U[\01^k]$, produces $\kbra{u_{i\circ a}}$ and then erases $i$.
The measurement \set{P_a,I_{2^d}-P_a} can also be performed efficiently (as any explicit measurement on \asO{\log n} qubits), the simplest way to do so is to represent the measurement as a projection in $\CC^{2^{d+1}}$ (recall that $d\in\asO{\log(n)}$) and perform that, using an auxiliary space of dimension $2^d$.

To see that $\ECm$ is a valid fingerprinting scheme with high probability, we will use \lemref{l_max_ip} together with the following technical lemma.

\lem[l_max_oip]{For $0\le i<2^r$, let $M$ be any mapping from an \f i-tuple of unit vectors in \RR[2^d]\ to a unit vector in \RR[2^d].
Then for any $s\in\01^{n-r}$, $\delta>0$, and $\set[a\in\01^n]{\ket{u_a}}$ defined over a randomly chosen $(n,r,{2^d})$-quasi-linear code $C$, it holds that $\sz{\bket{M(u_{0\circ s}\dc u_{(i-1)\circ s})}{u_{i\circ s}}}<\delta$ with probability at least $1-2\exp(-\delta^22^{d-1})$.}

\prf{Note that by the construction of quasi-linear codes, $\ket{u_{i\circ s}}$ is a uniformly random element of \set[\beta_1\dc\beta_{2^d}\in\{-1,1\}]{2^{-d/2}\sum_k\beta_k\ket{k}}, even if conditioned upon $v\deq M(u_{0\circ s}\dc u_{(i-1)\circ s})$.
So,
\m{\PR[C]{\sz{\bket{M(u_{0\circ s}\dc u_{(i-1)\circ s})}{u_{i\circ s}}}<\delta}
=\PR[C]{\sz{\sum_{k\in[2^d]}\beta_k v_k}<2^{d/2}\delta}
\ge1-2\exp(-2^{d-1}\delta^2),}
where the inequality follows from the Hoeffding bound (\lemref{l_Hoef}) and the fact that $\norm{v}=1$.}

Let us see that $\ECm$ is likely to be a valid fingerprinting scheme.

\lem[l_ECm_fp]{For $\ECm$ defined over a randomly chosen $(n+k,r,{2^d})$-quasi-linear code $C$, it holds that $\eps_-=0$ with certainty and $\eps_+<\delta$ with probability higher than $1-3\exp(n+r+k-\delta^22^{d-4k-7})$, for any $\delta>0$.}

\prf{Clearly, when $a_1=a_2$ the answer is always correct, that is $\eps_-=0$.

When, on the other hand, $a_1\ne a_2$, the probability of the wrong answer is $\tr(P_{a_2}\rho_{a_1})$.
Let $P_{a_2}'\deq\sum_{i\in\01^k}u_{i\circ a_2}u_{i\circ a_2}^*$; we will see that, with high probability over $C\sim\U[C]$, both $\tr(P_{a_2}'\rho_{a_1})$ and $\sz{\tr((P_{a_2}-P_{a_2}')\rho_{a_1})}$ are small.

\m[ECm1]{\tr\l(P_{a_2}'\rho_{a_1}\r)
=\sum_{i\in\01^k}\tr
\l(u_{i\circ a_2}u_{i\circ a_2}^*\rho_{a_1}\r)
\le2^k\delta_C^2,}
where $\delta_C\deq\Max[x_1\ne x_2]{\sz{u_{x_1}^*u_{x_2}}}$.

Observe that $P_{a_2}=\sum_{i\in\01^k}v_iv_i^*$, where \pl[v_i] are ``orthonormalized \pl[u_{i\circ a_2}]'', as follows
\m{v_0'=v_0\deq u_{0\circ a_2};\tb
v_i'\deq u_{i\circ a_2}
-\sum_{j<i}v_jv_j^*u_{i\circ a_2};\tb
v_i\deq v_i'/\sz{v_i'}.}
Let $\Delta_i\deq v_i-u_{i\circ a_2}$, then
\m{\sz{\Delta_i}\le\sz{u_{i\circ a_2}-v_i'}+\sz{v_i-v_i'}
\le2\sum_{j=0}^{i-1}\sz{v_j^*u_{i\circ a_2}}
\le2^k\Maxx{j}{\sz{v_j^*u_{i\circ a_2}}},}
and
\mal[ECm2]{\sz{\tr((P_{a_2}-P_{a_2}')\rho_{a_1})}
&\le\norm{P_{a_2}-P_{a_2}'}\\
&\le\sum_{i\in\01^k}
\norm{\l(u_{i\circ a_2}+\Delta_i)(u_{i\circ a_2}^*+\Delta_i^*\r)
-\l(u_{i\circ a_2}u_{i\circ a_2}^*\r)}\\
&\le3\tm2^k\Maxx{i}{\sz{\Delta_i}}
\le3\tm2^{2k}\Maxx{0\le j<i<2^k}{\sz{v_j^*u_{i\circ a_2}}}.}

Now we apply \lemref{l_max_oip}, where $M$ is the mapping that, according to our orthonormalization process, maps $(u_{k\circ a_2})_{k=0}^j$ to $v_j$.
For fixed $a_2$ and $j<i$, the lemma guarantees that $3\tm2^{2k}\sz{v_j^*u_{i\circ a_2}}$ is less than $\delta/2$ with probability at least $1-2\exp(-\delta^22^{d-4k-3}/9)$.
By the union bound, the right-hand side of \bref{ECm2} is less than $\delta/2$ with probability at least $1-2^{2k}\exp(-\delta^22^{d-4k-3}/9)>1-\exp(2k-\delta^22^{d-4k-7})$.
Another application of the union bound implies that the same holds for every $a_2$ with probability higher than $1-\exp(n+2k-\delta^22^{d-4k-7})$.

By \lemref{l_max_ip}, it holds that the right-hand side of \bref{ECm1} is less than $\delta/2$ (i.e., $2^k\delta_C^2<\delta/2$) with probability at least $1-2\exp(n+r+k-\delta2^{d-k})$.
Therefore, $\tr(P_{a_2}\rho_{a_1})<\delta$ for every $a_1\neq a_2$ with probability higher than $1-3\exp(n+r+k-\delta^22^{d-4k-7})$, as required.}

Our next step is a statement analogous to \lemref{l_ECp_Eh} that would apply to $\ECm$.
As before, we let $\rho_a'=2^{d-n}\rho_a$ and $\mu_v(a)=\bra v\rho_a'\ket v$ for arbitrary $v\in\CC^{2^d}$.

\lem[l_ECm_Eh]{Let $v\in\CC^{2^d}$ be a unit vector and $a_0\in\01^n$ be fixed, and assume that $\ECm$ is defined over an $(n+k,r,{2^d})$-quasi-linear code $C$, where ${2^k}\in\asom{\log n}$ and $d\in\asO{\log(n)}$.
Then
\m{\E[{C\sim\U[C]}]
{\Max{0,\mu_{v}(a_0)\log\l({2^n}\mu_{v}(a_0)\r)}}
\in\asO{\fr1{2^{n+k\l(\fr12-\lambda\r)}}}}
for every $\lambda>0$.
}

We will follow in the footsteps of our proof of \lemref{l_ECp_Eh}, however we will have to use somewhat ``heavier'' concentration tools.

\prf{For every $j\in\01^k$, let 
\m{\omega_v^a(j)\deq
\sz{\sum_{i\in[2^d]}sg\l(\ip{x^{(2)}}{c_i}
\xor {d_{x^{(1)}}}_i\r)v_i}^2,}
where $x=j\circ a$.
Then $\mu_{v}(a_0)=\fr1{2^{n+k}}\sum_{j\in\01^k}\omega_{v}^{a_0}(j)$.

For every $j$,
\m{\E[{C\sim\U[C]}]{\omega_{v}^{a_0}(j)}
=\E[{\beta_1\dc\beta_{2^d}\sim\U[\{-1,1\}]}]
{\sum_{i,j}\beta_i\beta_j{v}_i{v}_j}
=\norm{v}^2=1}
and $\E{\mu_{v}(a_0)}=1/{2^n}$.
Moreover, as we've seen in the proof of \lemref{l_ECp_Eh}, from \lemref{l_Hoef_c} and from $\norm{v}=1$ it follows that that for every $t\ge0$, $\PR{\omega_{v}^{a_0}(j)\ge t}\le4\exp(-t/4)$.
Therefore, by \theoref{t_Cher+} it holds that
\m{\PR[C]{\mu_{v}(a_0)\ge\fr{1+t}{2^n}}
\le\exp\l(\fr{-{2^k}t^2}{3904\l(\log\fr{16}t\r)^2}\r)\deq f(t)}
for $0<t\le4/7$.
Besides, it holds that $0\le\omega_{v}^{a_0}(j)\le2^d$.

As before, we define $g(x)\deq\Max{0,x\log(x)}$ and let $\tilde\mu$ be a new random variable that will replace $\mu_{v}(a_0)$ in further analysis.
We define the distribution of $\tilde\mu$ by demanding that $\PR{\tilde\mu\ge1+t}=f(t)$ for $0<t\le4/7$ and $\PR{\tilde\mu=2^d}=f(4/7)$.
The requirements of \lemref{l_E_subst} are satisfied by $g(\dt)$, $\mu$ and $\tilde\mu$, and therefore
\m{\E{\Max{0,\mu_{v}(a_0)\log\l({2^n}\mu_{v}(a_0)\r)}}
\le\fr1{2^n}\E{g(\tilde\mu)}.}

By the definition,
\m{\E{g(\tilde\mu)}
=\int_0^{4/7}(1+x)\log(1+x)\l(-\fr{df}{dx}\r)dx
+2^dd\tm f(4/7).}
Clearly, $f(4/7)\in\exp\l(-\asOm{2^k}\r)$ and $(1+x)\log(1+x)\l(-\fr{df}{dx}\r)\le {2^k}x^2f(x)$.
For every $\lambda>0$ there exists $A_{\lambda}>0$, such that $f(x)\le\exp\l(-A_{\lambda}{2^k}x^{2+\lambda}\r)$ for $0<x\le4/7$.
So,
\mal{\E{g(\tilde\mu)}
&<\int_0^{\infty}
{2^k}x^2\exp\l(-A_{\lambda}{2^k}x^{2+\lambda}\r)dx
+\exp\l(d+\log d-\asOm{2^k}\r)\\
&\le\fr{2^k}{(2+\lambda)\l(A_{\lambda}{2^k}\r)^{\fr3{2+\lambda}}}
\tm\Gamma\l(\fr3{2+\lambda}\r)
+\exp(d-\asOm{2^k}),}
where $\Gamma(a)\deq\int_0^{\infty}x^{a-1}\exp(-x)\,dx$ is the Gamma-function.
Therefore for ${2^k}\in\asom{\log n}$ and every $\lambda>0$,
\m{\E{g(\tilde\mu)}\le\asO{\fr1{2^{k\l(\fr12-\lambda\r)}}},}
as required.}

\ssect[s_fa]{Further security analysis of $\ECp$ and $\ECm$}

Based on \lemref[l_ECp_Eh]{l_ECm_Eh}, we continue our analysis of $\ECp$ and $\ECm$.
From this point on and unless stated otherwise, we view the former as a special case of the latter, corresponding to $k=0$.

First, as promised earlier, we prove that for almost all quasi-linear codes~$C$,
we have $\sum_a \rho'_a=I_{2^d}$.

\lem[l_avgstate]{If $C$ is an $(n+k,r,2^d)$-quasi-linear code
such that the vectors~$c_1,\dots,c_{2^d}$ are all distinct,
then $\sum_a \rho'_a=I_{2^d}$.
In particular, if an $(n+k,r,2^d)$-quasi-linear code~$C$ is chosen uniformly at random,
then $\sum_a \rho'_a=I_{2^d}$ with probability at least~$1-2^{2d+r-n-k}$.}

\prf{If $c_1,\dots,c_{2^d}$ are all distinct, then
\begin{align*}
\sum_a \rho'_a
&= 2^{d-n-k} \sum_{x\in\01^{n+k}} \kbra{u_x} \\
&= 2^{-n-k} \sum_{x^{(1)}} \sum_{i,j}
\sg\l((d_{x^{(1)}})_i\oplus(d_{x^{(1)}})_j\r) 
\left( \sum_{x^{(2)}} 
\sg\langle c_i\oplus c_j,x^{(2)}\rangle \right)
\kbra[i]{j}  %
= I_{2^d},
\end{align*}
where $x^{(1)}\in\01^r$, $x^{(2)}\in\01^{n+k-r}$, and $i,j\in[2^d]$.

Now let $C\sim\U[C]$.
For any fixed distinct $i$ and $j$, $c_i$ equals $c_j$ with probability~$2^{r-n-k}$.
By the union bound, the probability that all \pl[c_i] are distinct is at least
\m{1-\chs{2^d}2\tm2^{r-n-k}<2^{2d+r-n-k},}
as desired.}

Next we will argue that $\sum_{a\in\01^n}\mu_{v}(a)\log\l({2^n}\mu_{v}(a)\r)$ is unlikely to be large when $C\sim\U[C]$.

\lem[l_Ph]{Let $v\in\CC^{2^d}$ be a unit vector and assume that $C$ is a uniformly random $(n+k,r,{2^d})$-quasi-linear code, then for every $\delta>0$
\m{\PR[C]{\sum_{a\in\01^n}\mu_{v}(a)\log\l({2^n}\mu_{v}(a)\r)
>\alpha_k+\delta}
<\exp\l(n-2^{r-k-2d}\l(\fr{\delta}d\r)^2\r),}
where $\alpha_0<23$, and $\alpha_k\in\asO{1/2^{k(1/2-\lambda)}}$ for ${2^k}\in\asom{\log n}$ and any $\lambda>0$.}

\prf{We will use concentration bounds in conjunction with the mean guarantees of \lemref[l_ECp_Eh]{l_ECm_Eh}.

Define new random variables
\m{\tilde\mu(a)
\deq\Max{0,\mu_{v}(a)\log\l({2^n}\mu_{v}(a)\r)},}
then $0\le\tilde\mu(a)\le2^{d-n}d$.
From \lemref[l_ECp_Eh]{l_ECm_Eh} we know that
$\E[C]{\tilde\mu(a)}<23/{2^n}$ for $k=0$ and every $\lambda>0$, and
$\E[C]{\tilde\mu(a)}\in\asO{1/2^{n+k\l(1/2-\lambda\r)}}$
for ${2^k}\in\asom{\log n}$.

We want to bound the probability that $\sum_a\tilde\mu(a)>\delta$.
Let $t\deq r-k$, assume \wlg that $t>0$ and define
\m{A_i\deq\set[j\in\01^t]{j\circ i}}
for every $i\in\01^{n-t}$.
Observe that for every $i_0\in\01^n$ the random values \lrp[a\in A_{i_0}]{C(a)} are distributed identically and independently when $C\sim\U[C]$, and the same is true for \lrp[a\in A_{i_0}]{\tilde\mu(a)}.
Therefore the Hoeffding bound (\lemref{l_Hoef}) can be applied, resulting in
\m{\PR[U]{\sum_{a\in A_{i_0}}\tilde\mu(a)
>\fr{{2^n}\mu_0+\delta}{2^{n-t}}}
<2\exp\l(\fr{-2^{t+1}\delta^2}{2^{2d}d^2}\r),}
where $\mu_0\deq\E[C]{\tilde\mu(a)}$.
Therefore, from the union bound:
\m{\PR[U]{\sum_{a\in\01^n}\tilde\mu(a)>\alpha_k+\delta}
<2^{n-t+1}\exp\l(-2^{r-k-2d}\l(\fr{\delta}d\r)^2\r),}
as required.}

As we discussed before, if $\sum_{a\in\01^n}\mu_{v}(a)\log\l({2^n}\mu_{v}(a)\r)$ is small for a fixed $v$, that means that, informally, a holder of $\rho_a$ who has measured it and got the outcome $\kbra{v}$ has not learnt much about $a$.

Our next step will be to argue that, with high probability, $\sum_{a\in\01^n}\mu_{v}(a)\log\l({2^n}\mu_{v}(a)\r)$ is small for every pure state~$\ket{v}\in\CC[2^d]$.
According to the same intuition (which will be formalized soon), that would imply that no outcome of a measurement of $\rho_a$ exists, that can tell much about $a$.

First we argue that
the function~$\kbra{v}\mapsto\sum_{a\in\01^n}\mu_{v}(a)\log\l({2^n}\mu_{v}(a)\r)$ has a good continuity property (called the ``almost Lipschitz continuity'')
in order to discretize ``every pure state~$\ket{v}\in\CC[2^d]$'' in the above argument.

\lem[l_lipschitz]{Let~$C$ be an $(n+k,r,2^d)$-quasi-linear code, such that~$\sum_a\rho'_a=I_{2^d}$.
Let~$0<\eps\le2/e$ and $\ket{v}$ and~$\ket{w}$ be unit vectors in~$\CC^{2^d}$ such that $\norm{\kbra{v}-\kbra{w}}_1\le\eps$.
Then,
\[
\abs*{\sum_a\mu_v(a)\log(2^n\mu_v(a))-\sum_a\mu_w(a)\log(2^n\mu_w(a))}
\le 2^{d-1}\varepsilon\log\frac{2}{\varepsilon}.
\]}

\prf{Fix any~$a$ and we will prove $\abs{\mu_v(a)\log(2^{n-d}\mu_v(a))-\mu_w(a)\log(2^{n-d}\mu_w(a))}\le2^{d-n-1}\varepsilon\log(2/\varepsilon)$.
Without loss of generality, we can assume that~$\mu_v(a)\le\mu_w(a)$.
Then,
\[
\mu_w(a)-\mu_v(a)=2^{d-n}\tr\l(\rho_a(\kbra{w}-\kbra{v})\r)\le2^{d-n-1}\norm{\kbra{v}-\kbra{w}}_1\le2^{d-n-1}\varepsilon.
\]
Therefore,
\begin{align*}
&
\mu_w(a)\log(2^{n-d}\mu_w(a))-\mu_v(a)\log(2^{n-d}\mu_v(a)) \\
&=
\mu_w(a)\log(2^{n-d}\mu_w(a))-\mu_v(a)\log(2^{n-d}\mu_w(a))+\mu_v(a)\log(2^{n-d}\mu_w(a))-\mu_v(a)\log(2^{n-d}\mu_v(a)) \\
&=
(\mu_w(a)-\mu_v(a))\log(2^{n-d}\mu_w(a))+\mu_v(a)\log\left(1+\frac{\mu_w(a)-\mu_v(a)}{\mu_v(a)}\right).
\end{align*}
Note that $(\mu_w(a)-\mu_v(a))\log(2^{n-d}\mu_w(a))\le0$ and $\mu_v(a)\log(1+(\mu_w(a)-\mu_v(a))/\mu_v(a))\ge0$.
Therefore,
\begin{align*}
&
\abs{\mu_v(a)\log(2^{n-d}\mu_v(a))-\mu_w(a)\log(2^{n-d}\mu_w(a))} \\
&=
\abs*{(\mu_w(a)-\mu_v(a))\log(2^{n-d}\mu_w(a))+\mu_v(a)\log\left(1+\frac{\mu_w(a)-\mu_v(a)}{\mu_v(a)}\right)} \\
&\le
\max\left\{-(\mu_w(a)-\mu_v(a))\log(2^{n-d}\mu_w(a)),
\mu_v(a)\log\left(1+\frac{\mu_w(a)-\mu_v(a)}{\mu_v(a)}\right)\right\} \\
&\le
\max\left\{-(\mu_w(a)-\mu_v(a))\log(2^{n-d}(\mu_w(a)-\mu_v(a))),
\mu_v(a)\cdot\frac{\mu_w(a)-\mu_v(a)}{\mu_v(a)}\right\} \\
&\le
\max\left\{2^{d-n-1}\varepsilon\log\frac{2}{\varepsilon},2^{d-n-1}\varepsilon\right\} \\
&=
2^{d-n-1}\varepsilon\log\frac{2}{\varepsilon}.
\end{align*}
By the triangle inequality, we have
\[
\abs*{\sum_a\mu_v(a)\log(2^{n-d}\mu_v(a))-\sum_a\mu_w(a)\log(2^{n-d}\mu_w(a))}
\le
2^{d-1}\varepsilon\log\frac{2}{\varepsilon}.
\]
The left-hand side can be rewritten as
\begin{align*}
&
\abs*{\sum_a\mu_v(a)\log(2^{n-d}\mu_v(a))-\sum_a\mu_w(a)\log(2^{n-d}\mu_w(a))} \\
&=
\abs*{\sum_a\mu_v(a)\log(2^n\mu_v(a))-\sum_a\mu_w(a)\log(2^n\mu_w(a))+\left(\sum_a\mu_v(a)-\sum_a\mu_w(a)\right)\log2^{-d}} \\
&=
\abs*{\sum_a\mu_v(a)\log(2^n\mu_v(a))-\sum_a\mu_w(a)\log(2^n\mu_w(a))},
\end{align*}
which completes the proof.
}

We are ready to see that with high probability, $\sum\mu_{v}(a)\log\l({2^n}\mu_{v}(a)\r)$ is small for every $\ket{v}$.

\lem[l_maxPh]{Let~$C$ be a uniformly random $(n+k,r,2^d)$-quasi-linear code.
Let~$\delta>0$ satisfy that $e^{3/2}\delta/4\le2^d$.
Then,
\m{\PR[C]{\exists \ket{v}\colon
\sum_{a\in\01^n}\mu_v(a)\log\l({2^n}\mu_v(a)\r)
>\alpha_k+\delta}
<\exp\left(
2^{d+1}\log\frac{2^{2d+5}}{e^2\delta^2}
+n-2^{r-k-2d}\l(\fr{\delta}{2d}\r)^2
\right),}
where $\alpha_k$ is as in \lemref{l_Ph}.}

\prf{Let~$\varepsilon=2^{-2d-3}e^2\delta^2$.
By the assumption, we have~$\varepsilon\le2/e$.
Then we have
\[
2^{d-1}\varepsilon\log\frac{2}{\varepsilon}
=
\frac{e\delta}{2}\cdot\frac{e\delta}{2^{d+2}}\log\frac{2^{d+2}}{e\delta}
\le
\frac{e\delta}{2}\cdot\frac{1}{e}
=
\frac{\delta}{2},
\]
where the inequality follows from~$x\log(1/x)\le1/e$.
By \lemref{l_eps-net},
there exists an~$\varepsilon$-net~$M$ for the set of $2^d$-dimensional states
with respect to the trace distance
with size
\[
\abs{M}\le\left(\frac{4}{\varepsilon}\right)^{2^{d+1}}
=\left(\frac{2^{2d+5}}{e^2\delta^2}\right)^{2^{d+1}}.
\]
Suppose that the quasi-linear code~$C$ is such that
there exists a unit vector~$v$ such that
\[
\sum_{a\in\01^n}\mu_v(a)\log\l({2^n}\mu_v(a)\r)
>\alpha_k+\delta.
\]
Let~$w\in M$ be a unit vector satisfying~$\norm{\kbra{v}-\kbra{w}}_1\le\varepsilon$.
By \lemref{l_lipschitz},
\[
\sum_{a\in\01^n}\mu_w(a)\log(2^n\mu_w(a))
\ge
\sum_{a\in\01^n}\mu_v(a)\log(2^n\mu_v(a))
-
2^{d-1}\varepsilon\log\frac{2}{\varepsilon}
>
\alpha_k+\frac{\delta}{2}.
\]
This implies that
\begin{multline*}
\PR[C]{\exists \ket{v}\colon
\sum_{a\in\01^n}\mu_v(a)\log\l({2^n}\mu_v(a)\r)
>\alpha_k+\delta}
\\
\le
\PR[C]{\exists \ket{w}\in M\colon
\sum_{a\in\01^n}\mu_w(a)\log\l({2^n}\mu_w(a)\r)
>\alpha_k+\frac{\delta}{2}}.
\end{multline*}
By \lemref{l_Ph} and union bound,
the right-hand side is at most
\[
\abs{M}\cdot\exp\l(n-2^{r-k-2d}\l(\fr{\delta}{2d}\r)^2\r)
\le
\exp\left(
2^{d+1}\log\frac{2^{2d+5}}{e^2\delta^2}
+n-2^{r-k-2d}\l(\fr{\delta}{2d}\r)^2
\right),
\]
as required.
}

It remains to be seen that small values of $\sum\mu_{v}(a)\log\l({2^n}\mu_{v}(a)\r)$ for all $\ket{v}\in\CC[2^d]$ indeed imply good hiding properties of the corresponding fingerprinting scheme.

\lem[l_iacc]{Let~$C$ be an $(n+k,r,2^d)$-quasi-linear code
such that~$c_1,\dots,c_{2^d}$ are all distinct.
If~$a\in\01^n$ is chosen uniformly at random,
then the accessible information of the ensemble~$(\rho_a)$
is at most
\[
\max_{\ket{v}} \sum_{a\in\01^n}\mu_v(a)\log\l({2^n}\mu_v(a)\r).
\]}

\prf{We follow a similar path to that used in a proof in Section~2.2 of Leung~\cite{L09_A_Sur}.
Since the accessible information can be always achieved by a rank-one POVM,
let $M=\{\alpha_j \kbra{v_j}\}_j$ be a rank-one POVM achieving the accessible information,
where $\ket{v_j}$ is a pure state, $\alpha_j>0$ and $\sum_j\alpha_j=2^d$.
If $A$ is the random variable representing the choice of~$a$
and $J$ is the random variable representing the measurement result
of the state under~$M$,
then
\begin{align*}
\Iacc
&=H(J)-H(J\mid A) \\
&=-\sum_j \alpha_j \bra{v_j} \frac{I_{2^d}}{2^d} \ket{v_j} \log (\alpha_j \bra{v_j} \frac{I_{2^d}}{2^d} \ket{v_j})
+\frac{1}{2^n}\sum_{a,j}\alpha_j \bra{v_j} \rho_a \ket{v_j}
\log(\alpha_j \bra{v_j} \rho_a \ket{v_j}) \\
&=-\sum_j \frac{\alpha_j}{2^d} \log \frac{\alpha_j}{2^d}
+\frac{1}{2^n}\sum_{a,j}\alpha_j \bra{v_j} \rho_a \ket{v_j} \log\alpha_j
+\frac{1}{2^n}\sum_{a,j}\alpha_j \bra{v_j} \rho_a \ket{v_j}
\log \bra{v_j} \rho_a \ket{v_j} \\
&=-\sum_j \frac{\alpha_j}{2^d} \log \frac{\alpha_j}{2^d}
+\sum_j \frac{\alpha_j}{2^d} \log\alpha_j
+\frac{1}{2^n}\sum_{a,j}\alpha_j \bra{v_j} \rho_a \ket{v_j}
\log \bra{v_j} \rho_a \ket{v_j} \\
&=d\log 2
+\frac{1}{2^n}\sum_{a,j}\alpha_j \bra{v_j} \rho_a \ket{v_j}
\log \bra{v_j} \rho_a \ket{v_j} \\
&=d\log 2
+\frac{1}{2^n}\sum_{a,j}\alpha_j 2^{n-d} \bra{v_j} \rho'_a \ket{v_j}
\log(2^{n-d} \bra{v_j} \rho'_a \ket{v_j}) \\
&=\sum_{a,j}\frac{\alpha_j}{2^d} \mu_{v_j}(a) \log(2^n \mu_{v_j}(a)) \\
&\le\max_{\ket{v}} \sum_a \mu_v(a) \log(2^n \mu_v(a)),
\end{align*}
where the inequality follows from the convexity argument
(the convex combination is at most the maximum).}

Lemmas~\ref{l_ECp_fp}, \ref{l_ECm_fp}, \ref{l_avgstate}, \ref{l_maxPh} and~\ref{l_iacc} imply the following theorem:

\theo[t_Main]{For any constant $c$ there exist quantum fingerprinting schemes that
\itemi{
\item map \f n-bit strings to mixed states over \asO{\log n} qubits and whose error probability and accessible information are both bounded by $1/n^c$;
\item map \f n-bit strings to pure states over \asO{\log n} qubits, whose error probability is bounded by $1/n^c$ and accessible information is \asO1.
}
The schemes are computationally efficient and have one-sided error with $\eps_-=0$ (answers ``$x\neq y$'' are always true).
}

\prf{Let $k=\ceil{4c\lg n}$, $d=\ceil{(18c+1)\lg n}$ and $r=\ceil{(60c+3)\lg n}$,
and let $\ECm$ be the mixed-state fingerprinting scheme defined over a randomly chosen $(n+k,r,2^d)$-quasi-linear code~$C$.
By Lemma~\ref{l_ECm_fp}, the probability that $\varepsilon_+\ge1/n^c$ vanishes as $n\to\infty$.

The probability that $C$ violates the condition of \lemref{l_avgstate} is negligible, so we assume the opposite, that allows us to use \lemref{l_iacc}.
Applying \lemref{l_maxPh} with~$\delta=1/(2n^c)$ to Lemma~\ref{l_iacc} and noting that $\alpha_k\in\asO{1/2^{k/3}}\subseteq\aso{1/n^c}$,
we obtain that the accessible information is at most $1/n^c$. 

Choosing $k=0$ and adjusting $d$ and $r$ accordingly gives the desired result for $\ECp$.}

Note that only polynomial amount of randomness is required in order to describe any of our fingerprinting schemes.
Moreover, a random string may be published openly without compromising the hiding guarantees of the schemes.

Mixed-state schemes can be viewed as a natural generalization of pure-state ones.
Our mixed-state construction achieves much better hiding guarantees (in the following section we argue its optimality), but even the pure-state one already reaches beyond the limitations of classical schemes, where we've seen~(cf.~\sref{ss_claqua_hid}) that $\asOm{\log(1/\eps)}$ bits are leaked by any scheme with error at most $\eps$.

\sect[s_opti]{Optimality of our schemes}

In this part we construct a generic strategy for extracting information from arbitrary quantum fingerprints.
We give a strategy that retrieves at least $1/\poly(D)$ bits of information about $x$ from a (w.l.g., mixed-state) fingerprint of $x$ over $\log D$ qubits.

We note that the following ``no-go'' argument remains valid for some weaker versions of fingerprinting than what is guaranteed by \theoref{t_Main}, namely:\itemi{
\item schemes with two-sided error;
\item schemes that only work in average \wrt ``balanced uniform'' input distribution (i.e., $(x,y)\sim(\U[A]+\U[B])/2$, where $A=\set{(x,x)}$ and $B=\set[x\neq y]{(x,y)}$).
}

To extract classical information about unknown $x\sim\U[\01^n]$ from its fingerprint $\rho(x)\in\CC[D\times D]$, we apply to $\rho(x)$ a complete projective measurement
\m{P_V\deq\set[v\in V]{\kbra{v}},}
where $V$ is a uniformly chosen random orthonormal basis for $\CC[D]$.\fn
{The idea of using randomly chosen projective measurements in order to prove a lower bound on accessible information has appeared in~\cite{JRW94_Low}.
However, our setting and the analysis are different.}
We will see that the mutual information between the outcome of $P_V$ and $x$ is at least $1/\poly(D)$.

\ssect[ss_opti_tech]{Technical preliminaries}

Optimality of our scheme from \sref{s_schemes} will follow from several technical lemmas that we state next.

It is well known that the ``distinguishability'' of two arbitrary quantum states $\sigma_1$ and $\sigma_2$ is determined by their trace distance $\norm[1]{\sigma_1-\sigma_2}$.
Informally speaking, we will show that \e{a randomly chosen complete projective measurement distinguishes between $\sigma_1$ and $\sigma_2$ only $\poly(D)$ times less efficiently than a best distinguishing measurement}.

Let $\Uu[1]{D}$ denote the uniform distribution of unit vectors in \CC[D].
The following is a well-known fact about $\Uu[1]{D}$.

\clm[c_uni]{Sampling $v\sim\Uu[1]{D}$ can be realized via the following algorithm:\itemi{
\item[1] Independently sample $u_r^1\dc u_r^D$ and $u_i^1\dc u_i^D$ from the standard normal distribution $N(0,1)$.
\item[2] Let $v\deq u/\norm u$ where $u\deq\l(u_r^j+u_i^j\tm\ii\r)_{j=1}^D$.
}}

\prf{The density function of $u$ is spherically symmetric.}

We need several technical lemmas.
First, let us see that the length of the projection of a randomly chosen vector $v\sim\Uu[1]{D}$ to any subspace cannot be ``too concentrated'':

\lemdup{l_devi}{\lemDevi}{Let $A\sbs[D]$, $1\le\sz A<D$.
Then for some $\eta_1\in\asOm{\fr1{D^2\log D}}$ and $\eta_2\in\asOm{\fr1{D^2(\log D)^4}}$,
\m{\PR[{v\sim\Uu[1]{D}}]
{\sum_{i\in A}\sz{v^i}^2\ge\fr{\sz{A}}D+\eta_1}\ge\eta_2.}
}

It is easy to see (by linearity of expectation and the fact that $\sz v=1$) that $\E[v]{\sum_A\sz{v^i}^2}=\sz{A}/D$, and therefore the above statement can be viewed as complementary to concentration bounds.

\prf{In the notation of \clmref{c_uni},
\mal[m_Opt3]{\PR[{v\sim\Uu[1]{D}}]
{\sum_{i\in A}\sz{v^i}^2\ge\fr{\sz{A}}D+\eps}
&=\PR{\sum_{i\nin A}\sz{v^i}^2\le1-\fr{\sz{A}}D-\eps}\\
&=\PR{\fr{\sum_{i\in A}\sz{v^i}^2}
{\sum_{i\nin A}\sz{v^i}^2}\ge
\fr{\sz{A}+D\eps}{D-\sz{A}-D\eps}}\\
&=\PR{\fr{\sum_{i\in A}((u_r^j)^2+(u_i^j)^2)}
{\sum_{i\nin A}((u_r^j)^2+(u_i^j)^2)}
\ge\fr{\sz{A}+D\eps}{D-\sz{A}-D\eps}}\\
&\ge\PR{Y^+\ge2\sz{A}+2D\eps}
\tm\PR{Y^-\le2D-2\sz{A}-2D\eps},}
where $Y^+\deq\sum_{i\in A}((u_r^j)^2+(u_i^j)^2)$, $Y^-\deq\sum_{i\nin A}((u_r^j)^2+(u_i^j)^2)$, and the inequality follows from $Y^+$ and $Y^-$ being mutually independent.

We analyze the behavior of $Y^+$ and $Y^-$.
Let ``$\odot$'' stand for either ``$+$'' or ``$-$''.
The distribution of $Y^\odot$ is known as $\chi_{k^\odot}^2$, where $k^+\deq2\sz{A}$ and $k^-\deq2D-2\sz{A}$; its density function is 
\m{\psi^\odot(x)=\fr1{2^{k^\odot/2}\Gamma(k^\odot/2)}\exp\l(-\fr x2\r)x^{k^\odot/2-1}} (cf.~\cite{JKB94_Con_U}).
One can see that $\E{Y^\odot}=k^\odot$ and $\E{\l(Y^\odot\r)^2}={k^\odot}^2+2k^\odot$ (thus, $\Var{Y^\odot}=2k^\odot$).

For $\gamma^\odot\deq5k^\odot\log(k^\odot)+20$, let $Y_{\gamma^\odot}^\odot$ be distributed as $Y^\odot$ modulo $Y^\odot\le\gamma^\odot$.
The density function of $Y_{\gamma^\odot}^\odot$ is
\m{\psi_{\gamma^\odot}^\odot(x)
=\twocase{\alpha_{\gamma^\odot}\psi(x)}{if $x\le\gamma^\odot$}{0}{else},}
for $\alpha_{\gamma^\odot}\deq1/\PR{Y^\odot\le\gamma^\odot}$.
Then
\m{k^\odot\ge\E{Y_{\gamma^\odot}^\odot}
=\alpha_{\gamma^\odot}
\l(k^\odot-\int_{\gamma^\odot}^{\infty}x\psi^\odot(x)\,dx\r)
\ge k^\odot-\zeta^\odot}
and
\m{\E{\l(Y_{\gamma^\odot}^\odot\r)^2}
=\alpha_{\gamma^\odot}
\l({k^\odot}^2+2k^\odot
-\int_{\gamma^\odot}^{\infty}x^2\psi^\odot(x)\,dx\r)
\ge {k^\odot}^2+2k^\odot-\zeta^\odot,}
where
\m[m_Opt-zeta]{\zeta^\odot\deq
\int_{\gamma}^{\infty}x^2\psi^\odot(x)\,dx
\le\fr1{2^{k^\odot/2}\Gamma(k^\odot/2)}
\int_{\gamma^\odot}^\infty\exp\l(-\fr x4\r)dx}
(the inequality follows from $x^2\tm\exp(-x/2)x^{k^\odot/2-1}\le\exp(-x/4)$, as guaranteed by our choice of $\gamma^\odot$).
In particular, $\zeta^\odot<1$ and $\Var{Y_\gamma^\odot}\ge2k^\odot-\zeta^\odot>k^\odot$ and
\m[m_Opt4]{
\E{\sz{Y_{\gamma^\odot}^\odot-\E{Y_{\gamma^\odot}^\odot}}}
\ge\Var{Y_{\gamma^\odot}^\odot}/\gamma^\odot
>k^\odot/\gamma^\odot.}

Denote:
\mal{\mu^\odot&\deq\E{Y_{\gamma^\odot}^\odot}&\tbbb
\Delta^\odot&\deq\E{\sz{Y_{\gamma^\odot}^\odot-\mu^\odot}}\\
\mu_+^\odot&\deq
\Ee{Y_{\gamma^\odot}^\odot}{Y_{\gamma^\odot}^\odot\ge\mu^\odot}&
q_+^\odot&\deq\PR{Y_{\gamma^\odot}^\odot\ge\mu^\odot}\\
\mu_-^\odot&\deq
\Ee{Y_{\gamma^\odot}^\odot}{Y_{\gamma^\odot}^\odot<\mu^\odot}&
q_-^\odot&\deq\PR{Y_{\gamma^\odot}^\odot<\mu^\odot}}
Then
\mal{&q_+^\odot\mu_+^\odot+q_-^\odot\mu_-^\odot=\mu^\odot,\\
&q_+^\odot\l(\mu_+^\odot-\mu^\odot\r)
+q_-^\odot\l(\mu^\odot-\mu_-^\odot\r)=\Delta^\odot,\\
&q_+^\odot+q_-^\odot=1,}
which implies 
\m[m_Opt4.5]{q_+^\odot\l(\mu_+^\odot-\mu^\odot\r)
=q_-^\odot\l(\mu^\odot-\mu_-^\odot\r)=\Delta^\odot/2.}
Clearly, $0\le Y_{\gamma^\odot}^\odot\le\gamma^\odot$ implies that
\m{\PR{Y_{\gamma^\odot}^\odot\ge\mu_+^\odot-\beta}
>\fr{q_+^\odot\beta}{\gamma^\odot}
\txt{~~~and~~~}
\PR{Y_{\gamma^\odot}^\odot\le\mu_-^\odot+\beta}
>\fr{q_-^\odot\beta}{\gamma^\odot}}
for every $\beta>0$.
Choosing $\beta=(\mu_+^+-\mu^+)/2$ gives
\m{\PR{Y_{\gamma^+}^+\ge\fr[]{\l(\mu^++\mu_+^+\r)}2}
\ge\fr{q_+^+\l(\mu_+^+-\mu^+\r)}{2\gamma^+}
=\fr{\Delta^+}{4\gamma^+},}
and similarly, via $\beta=(\mu^--\mu_-^-)/2$ one obtains
\m{\PR{Y_{\gamma^+}^-\le\fr[]{\l(\mu^-+\mu_-^-\r)}2}
\ge\fr{\Delta^-}{4\gamma^-}.}

On the other hand, \bref{m_Opt4.5} implies that $\mu_+^+-\mu^+\ge\Delta^+/2$ and $\mu^--\mu_-^-\ge\Delta^-/2$.
Therefore, from \bref{m_Opt4}:
\m{\PR{Y_\gamma^+\ge k^+-\zeta^++\fr{k^+}{2\gamma^+}}
\ge\PR{Y_{\gamma^+}^+\ge\mu^++\Delta^+/2}
\ge\fr{\Delta^+}{4\gamma^+}\ge\fr{k^+}{4{\gamma^+}^2},}
and similarly,
\m{\PR{Y_\gamma^-\le k^--\fr{k^-}{2\gamma^-}}
\ge\fr{k^-}{4{\gamma^-}^2}.}
From \bref{m_Opt-zeta} it is obvious that $\zeta^+<\fr1{4\gamma^+}$, and therefore, by the definition of $Y_{\gamma^+}^+$,
\m{\PR{Y^+\ge2\sz{A}+\fr1{4\gamma^+}}
\ge\PR{Y_{\gamma^+}^+\ge k^++\fr1{4\gamma^+}}
\ge\fr{k^+}{4{\gamma^+}^2}.}
By the definition of $Y_{\gamma^-}^-$ and the obvious fact that $\PR{Y^-\le\gamma^-}>1/2$,
\m{\PR{Y^-\le2D-2\sz{A}-\fr1{2\gamma^-}}
\ge\PR{Y^-\le\gamma^-}
\tm\PR{Y_{\gamma^-}^-\le k^--\fr{k^-}{2\gamma^-}}
>\fr{k^-}{8{\gamma^-}^2}.}

Observe that $\fr{k^\odot}{{\gamma^\odot}^2}\ge\fr1{51D(\log D)^2}$ and $\fr1{\gamma^\odot}\ge\fr1{11D\log D}$ for large enough $D$.
Together with \bref{m_Opt3} this implies
\m{\PR[{v\sim\Uu[1]{D}}]
{\sum_{i\in A}\sz{v^i}^2\ge\fr{\sz{A}}D+\fr1{88D^2\log D}}
\ge\fr1{83232\tm D^2(\log D)^4},}
as required.
}

Denote by $\Ub$ the uniform distribution of orthonormal bases of \CC[D] (i.e., the Haar measure).
For $\rho\in\Den D$, we will write $\Prho$ to denote the distribution of the outcome of $P_V(\rho)$ when $V\sim\Ub$.
We will implicitly identify an outcome of $\Prho$ with the corresponding unit vector in \CC[D].

We need yet another ``anti-concentration'' statement, this time to say that the outcomes of $\Prho$ cannot be too concentrated for any fixed $\rho$:

\lemdup{l_uni}{\lemUni}{Let $B$ be a subset of unit vectors in \CC[D], such that $\Uu[1]{D}(B)\ge\eps$.
Then for any $\rho\in\Den D$,
\m{\PR[v\sim\Prho]{v\in B}>\fr{\eps^4}{256}.}
}

Intuitively, by choosing $\rho$ adversarially one can selectively ``hide'' some unit vectors in \CC[D] from $\Prho$.
However, only those \pl[v] are hidden well that are almost orthogonal to all spectral components of $\rho$, and that cannot happen to too many \pl[v] simultaneously; in particular, if $B$ is sufficiently large then it is impossible to efficiently avoid all its elements.

\prf{Observe that the distribution $\Uu[1]{D}$ is the same as $P_{V\sim\Ub}(I_D/D)$, and its density function is constant on the support (unit vectors in \CC[D]) -- denote it by $\phi_0$.
Then by linearity, for any $\rho$ the density function of $\Prho$ is
\m{\phi_\rho(v)\deq\phi_0\tm D\tm\bket[\rho]vv.}

For $\delta\deq\eps^3/64$, let us bound from above the value of
\m[m_uni1]{\PR[{v\sim\Uu[1]{D}}]{\phi_\rho(v)<\delta\tm\phi_0}
=\PR[{\Uu[1]{D}}]{\bket[\rho]vv<\delta/D}.}
The expectation of $\bket[\rho]vv$ is $1/D$, and therefore the value is maximized when $\rho$ has rank one (if $\rho$ is a mixture that makes the value of $\bket[\rho]vv$ more concentrated).
On the other hand, for every fixed $u_0$ and $v\sim\Uu[1]{D}$, the distribution of $\sz{\bket{u_0}v}$ only depends on $\sz{u_0}$ (and not on the ``direction'' of $u_0$).
Therefore, in order to bound \bref{m_uni1}, we can assume \wlg that $\rho=\kbra{u_0}$, where $u_0=(1,0\dc0)$.
That is,
\m{\PR[{v\sim\Uu[1]{D}}]{\phi_\rho(v)<\delta\tm\phi_0}
\le\PR[{\Uu[1]{D}}]{\sz{v^1}<\sqrt{\delta/D}},}
where $v^1$ is the first coordinate of $v$.

By \clmref{c_uni} we have:
\m{\PR[{v\sim\Uu[1]{D}}]{\sz{v^1}<\sqrt{\fr[]\delta D}}
=\PR{\fr[]{\sz{u^1}}{\norm{u}}<\sqrt{\fr[]\delta D}}
\le\PR{\sz{u^1}<2\sqrt{\delta/\eps}}
+\PR{\norm{u}^2>\fr{4D}\eps}.}
We know that $\norm{u}^2\sim\chi_{2D}^2$, and therefore its expectation is $2D$ and $\PR{\norm{u}^2>4D/\eps}<\eps/2$ by Markov inequality.
We also know that $\Re(u^1)\sim N(0,1)$, and therefore $\PR{\sz{u^1}<2\sqrt{\delta/\eps}}<2\sqrt{\delta/\eps}=\eps/4$.
We conclude that $\PR[{v\sim\Uu[1]{D}}]{\phi_\rho(v)<\delta\tm\phi_0}<3\eps/4$.

Let $B'\deq\set[\phi_\rho(v)\ge\delta\tm\phi_0]{v\in B}$, then it necessarily holds that $\Uu[1]{D}(B')>\eps/4$.
By the definition of $B'$,
\m{\PR[v\sim\Prho]{v\in B'}\ge\delta\tm\Uu[1]{D}(B')
>\fr{\delta\eps}4=\fr{\eps^4}{256},}
and the result follows.}

The next lemma will be the core of our optimality argument.

\lemdup{l_dist_ra}{\lemDisRa}{Let $\sigma_1,\sigma_2,\rho\in\Den D$, satisfying $\norm[1]{\sigma_1-\sigma_2}=\delta>0$.
Then for some $\xi\in\asOm{\fr{\delta}{D^3\log D}}$,
\m{\PR[v\sim\Prho]
{\bket[\sigma_1]{v}{v}\ge(1+\xi)\bket[\sigma_2]{v}{v}}
\in\asOm{(D\log D)^{-20}}.}
}

\prf{To prove the statement, we will first consider the simpler case when $v\sim\Uu[1]{D}$, then see what happens when $v\sim\Prho$.

Let $\sigma'\deq\sigma_1-\sigma_2$, then
\m{\PR[{v\sim\Uu[1]{D}}]{\bket[\sigma_1]vv\ge(1+\xi)\bket[\sigma_2]vv}
=\PR{\bket[\sigma']vv\ge\xi\bket[\sigma_2]vv}
\ge\PR{\bket[\sigma']vv\ge\xi}.}
Let $\sigma'=\sum_{i=1}^De_i\kbra{u_i}$ be a spectral decomposition, $A^+\deq\set[e_i>0]i$ and $A^-\deq\set[e_i<0]i$, then for every $\xi$
\mal[m_Opt1]{\PR[{v\sim\Uu[1]{D}}]{\bket[\sigma']vv\ge\xi}
&=\PR{\sum_ie_i\sz{\bket{u_i}{v}}^2\ge\xi}\\
&=\PR{\sum_{i\in A^+}e_i\sz{\bket{u_i}{v}}^2\ge\xi
+\sum_{i\in A^-}-e_i\sz{\bket{u_i}{v}}^2}\\
&\ge\PR{\sum_{i\in A^+}e_i\sz{\bket{u_i}{v}}^2\ge
\xi+\E[{v\sim\Uu[1]{D}}]
{\sum_{i\in A^+}e_i\sz{\bket{u_i}{v}}^2}},}
where the inequality follows from $\sum e_i=0$ and the fact that the random values $\sum_{A^+}e_i\sz{\bket{u_i}{v}}^2$ and $\sum_{A^-}-e_i\sz{\bket{u_i}{v}}^2$ are anti-correlated when $v\sim\Uu[1]{D}$.

Observe that $\sum\sz{e_i}=\delta$, and so $\sum_{A^+}e_i=\delta/2$.
As $\E[v]{\sz{\bket uv}^2}=1/D$ for any unit vector $u$ and the right-hand side of \bref{m_Opt1} is symmetric \wrt any unitary rotation of the vectors $\set{u_i}$,
\m[m_Opt2]{\PR[{v\sim\Uu[1]{D}}]{\bket[\sigma']vv\ge\xi}
\ge\PR{\sum_{i\in A^+}e_i\tm\sz{v^i}^2
\ge\xi+\fr{\delta}{2D}}.}

From \lemref{l_devi}, for some $\eta_1\in\asOm{\fr1{D^2\log D}}$ and $\eta_2\in\asOm{\fr1{D^2(\log D)^4}}$
\m{\PR[{v\sim\Uu[1]{D}}]
{\sum_{i\in A^+}\sz{v^i}^2\ge\fr{\sz{A^+}}D+\eta_1}\ge\eta_2.}
By the linearity of expectation,
\m{\Ee{\sum_{i\in A^+}e_i\tm\sz{v^i}^2}
{\sum_{i\in A^+}\sz{v^i}^2\ge\fr{\sz{A^+}}D+\eta_1}
\ge\fr{\delta}{2D}\tm\fr{\sz{A^+}+\eta_1D}{\sz{A^+}}
\ge\fr{\delta}{2D}+\fr{\delta\eta_1}{2D}.}
Therefore, for some $\xi\in\asOm{\fr{\delta}{D^3\log D}}$ and $\eta_3\in\asOm{\fr1{(D\log D)^5}}$,
\mal{\PR[{v\sim\Uu[1]{D}}]{\sum_{i\in A^+}e_i\tm\sz{v^i}^2
\ge\fr{\delta}{2D}+\xi}
&=\PR[{v\sim\Uu[1]{D}}]{\sum_{i\in A^+}e_i\tm\sz{v^i}^2
\ge\fr{\delta}{2D}+\fr{\delta\eta_1}{4D}}\\
&\ge\PR{\sum_{i\in A^+}\sz{v^i}^2\ge\fr{\sz{A^+}}D+\eta_1}
\tm\fr[b_]{\fr{\delta\eta_1}{4D}}{\sum_{A^+}e_i}\\
&\ge\fr{\eta_1\eta_2}{2D}=\eta_3.}
From \bref{m_Opt2}, $\PR[{v\sim\Uu[1]{D}}]{\bket[\sigma']vv\ge\xi}\ge\eta_3$.

Applying \lemref{l_uni} to the set $\set[{\bket[\sigma']vv\ge\xi},~\norm v=1]{v\in\CC[D]}$, we conclude that
\m{\PR[{v\sim\Prho}]{\bket[\sigma']vv\ge\xi}
\ge\fr{(\eta_3)^4}{256}
\in\asOm{\fr1{(D\log D)^{20}}},}
and the result follows.}

\ssect{Optimality statement}

The following theorem concludes our optimality argument.

\theo[t_optimal]{Let $\Phi=\set[x\in\01^n]{\phi(x)}\sbs\Den D$ be a quantum fingerprinting scheme that guarantees error below $1/2-\asOm1$.
Then $\Phi$ leaks \asOm{D^{-47}} bits of information.}

The theorem implies that any quantum fingerprinting scheme that leaks $\ell$ bits about $x$ requires \asOm{\log(1/\ell)} qubits, and therefore our mixed-state construction of \sref{s_mixed} (cf.~\theoref{t_Main}) is optimal.
Note that while our constructions of fingerprinting schemes guarantee one-sided error, the above theorem remains valid also for schemes with two-sided error.
Moreover, \theoref{t_optimal} theorem still holds for schemes that only work on average under the balanced uniform input distribution. 

\prf{We will show that for any $\Phi$, a measurement $P_V$ chosen at random \wrt $V\sim\Ub$ is likely to have the following property:
\e{The outcome of $P(\phi(X))$ has mutual information \asOm{D^{-47}} with the random variable $X\sim\U[\01^n]$.}

Assume $X=x_0$.
Let $\rho\deq\E[x\in\01^n]{\phi(x)}$.
Call a unit vector $v\in\CC[D]$ \e{$x_0$-$\eps$-good} if $\bket[\phi(x_0)]vv\ge(1+\eps)\bket[\rho]vv$, where $\eps\ge0$.

The error guarantee of the theorem implies that $\norm[1]{\phi(x_0)-\rho}\in\asOm1$ (as long as $n>0$), and therefore by \lemref{l_dist_ra},
\m[m_tOpt1]{\PR[v\sim\Prho]{v\txt{ is $x_0$-$\xi$-good}}
\in\asOm{(D\log D)^{-20}}}
for some $\xi\in\asOm{\fr[]1{D^3\log D}}$.

For any unit vector $v\in\CC[D]$, let $A_v$ be the set of all \pl[x], such that $v$ is $x$-$\xi$-good.
Let
\m{p_0\deq\PR[\mac{X\sim\U[\01^n]\\v\sim\Prho}]
{X\in A_v}\txt{~~~~and~~~~}
p_1\deq\PR[\mac{X\sim\U[\01^n]\\v\sim P_{V\sim\Ub}(\phi(X))}]
{X\in A_v}.}
By the definition of $x_0$-$\eps$-good we know that $p_1\ge(1+\xi)\tm p_0$. 

Note that $p_1$ is the ``actual'' probability of certain event (namely, $X\in A_v$), and $p_0$ is what that probability would have been if the outcome of $P_{V\sim\Ub}(\phi(X))$ did not depend on $X$.
Based on the inequality between the two probabilities, we want to show that the outcome of the measurement is \e{well-correlated} with the value of $X$.
For that we use a lower bound on $p_0$, as guaranteed by \bref{m_tOpt1}.

Now assume that the underlying distributions are $X\sim\U[\01^n]$ and $v\sim P_{V\sim\Ub}(\phi(X))$.
\m{\hh{X}v\le-p_1\tm\log_2\l(2^{-n}\tm\fr{p_1}{p_0}\r)
-(1-p_1)\tm\log_2\l(2^{-n}\tm\fr{1-p_1}{1-p_0}\r),}
as follows from the fact that the maximum entropy of a discrete distribution over a domain of given size is attained when the distribution is uniform (so, in the right-hand side we consider the situation when $X$ is uniform both modulo ``$X\in A_v$'' and modulo ``$X\nin A_v$'').
Then
\m{\hh{X}v\le n-p_1\log_2\l(\fr{p_1}{p_0}\r)
-(1-p_1)\log_2\l(\fr{1-p_1}{1-p_0}\r)=
n-\KL{D_0}{D_1},}
where $D_i$ is the distribution over \01 that assigns weight $p_i$ to the outcome ``0''.
By the Pinsker's inequality,
\m{\KL{D_0}{D_1}\ge\fr{\norm{D_0-D_1}_1^2}2
=2(p_1-p_0)^2\ge2(\xi p_0)^2\in\asOm{D^{-47}},}
and therefore
\m{\h{X}-\hh{X}v\in\asOm{D^{-47}}.}

Since $v$ is the outcome of a measurement performed on a fingerprint of $X$, the result follows.
}

\subsection*{Acknowledgments}
The authors express their appreciation to Debbie Leung, J{\'e}r{\'e}mie Roland, Martin R{\"o}tteler and Pranab Sen for valuable discussions.

\bibliography{tex}

\newpage
\appendix

\sect{Proof of \lemref{l_eps-net}}

Let us repeat the lemma:
\lemrep{l_eps-net}{\lemEpsNet}

To prove the lemma we use the following lemma that has been stated in~\cite{JRW94_Low}, where it was attributed to~\cite{S74_Qua_T}.

\nlem[l_uniform]{(\cite{JRW94_Low})}{
Let $\{\ket{e_1},\dots,\ket{e_D}\}$ be an orthonormal basis of $\CC^D$.
Let $\ket{u}\in\CC^D$ be a random unit vector
chosen according to the unitarily invariant probability distribution on the unit sphere in~$\CC^D$.
Let $X_i=\abs{\bket{e_i}{u}}^2$ for $i=1,\dots,D$.
Then, the range of the $D$-tuple $\vec{X}=(X_1,\dots,X_D)$
is equal to the probability simplex
\[
\Delta_{D-1}=\left\{(x_1,\dots,x_D)\colon
\sum_{i=1}^D x_i=1,\;
x_i\ge0\;(\forall i)\right\},
\]
and the probability distribution of $\vec{X}$ is uniform on $\Delta_{D-1}$.
}

\begin{corollary} \label{corollary:random-inner-product}
Let~$\ket{w}\in\CC^D$ be a fixed unit vector.
Choose a unit vector~$\ket{u}\in\CC^D$ randomly as in \lemref{l_uniform}.
Then $\PR{\abs{\bket{u}{w}}^2\ge x}=(1-x)^{D-1}$ for $0\le x\le1$.
\end{corollary}

\prf[\lemref{l_eps-net}]{
The lemma can be proved by the packing argument
in the same way as
\fakelemref{II.4} of~\cite{HLSW04_Ran_Qua}
and \fakelemref{4} of~\cite{BHLSW05_Rem_Pre}.
The difference is that we apply the packing argument
directly on the set of pure states
by using Corollary~\ref{corollary:random-inner-product},
instead of applying the packing argument
on the Euclidean space~$\RR^{2D}$ as an intermediate step.

Let $M$ be a maximal subset of~$\{\ket{v}\in\CC^D\colon\norm{v}=1\}$
such that every pair of distinct vectors~$\ket{u},\ket{v}\in M$
satisfy $\norm{\kbra{u}-\kbra{v}}_1\ge\varepsilon$.
By the maximality of $M$,
$M$ is an $\varepsilon$-net for the set of pure states in~$\CC^D$
with respect to the trace distance.
For each $\ket{u}\in M$, consider the open ball~%
$B_{\varepsilon/2}(\ket{u})=\{\ket{w}\in\CC^D\colon\norm{w}=1\wedge\norm{\kbra{u}-\kbra{w}}_1<\varepsilon/2\}$.
First fix $\ket{u}\in M$.
Then, if we pick a unit vector~$\ket{x}$ uniformly at random,
we have
\begin{align*}
\PR{\ket{x}\in B_{\varepsilon/2}(\ket{u})}
&=\PR{\norm{\kbra{u}-\kbra{x}}<\frac{\varepsilon}{2}} \\
&=\PR{\abs{\bket{u}{x}}^2>1-\left(\frac{\varepsilon}{4}\right)^2}
=\left(\frac{\varepsilon}{4}\right)^{2(D-1)},
\end{align*}
by Corollary~\ref{corollary:random-inner-product}.
By the condition of $M$,
the $\abs{M}$ open balls~$B_{\varepsilon/2}(\ket{u})$ ($\ket{u}\in M$) are disjoint.
Therefore,
\begin{align*}
1\ge\Pr\left[x\in\bigcup_{\ket{u}\in M}B_{\varepsilon/2}(\ket{u})\right]
=\sum_{\ket{u}\in M}\Pr[\ket{x}\in B_{\varepsilon/2}(\ket{u})]
=\abs{M}\left(\frac{\varepsilon}{4}\right)^{2(D-1)},
\end{align*}
which implies $\abs{M}\le(4/\varepsilon)^{2(D-1)}$.}

\end{document}